\documentclass[aps,prb,superscriptaddress,twocolumn,10pt,longbibliography]{revtex4-2}

\usepackage{amsmath}
\usepackage{graphicx}
\usepackage{bm}
\usepackage{color}

\date{\today}

\definecolor{darkblue}{rgb}{0.1,0.2,0.6}
\definecolor{darkred}{rgb}{0.8,0.1,0.2}
\usepackage[colorlinks,citecolor=darkblue,linkcolor=darkred,urlcolor=darkblue]{hyperref}
\usepackage[all]{hypcap}

\usepackage{etoolbox}
\usepackage{mathtools}
\usepackage{amsmath}
\usepackage{amsfonts}
\usepackage{amssymb}
\usepackage[normalem]{ulem}

\begin{document}

\title{Ancilla quantum measurements on interacting chains: Sensitivity of entanglement dynamics to the type and concentration of detectors}
\author{Elmer~V.~H.~Doggen}
\affiliation{\mbox{Institute for Quantum Materials and Technologies, Karlsruhe Institute of Technology, 76131 Karlsruhe, Germany}}
\affiliation{\mbox{Institut f\"ur Theorie der Kondensierten Materie, Karlsruhe Institute of Technology, 76131 Karlsruhe, Germany}}
\author{Igor V.~Gornyi}
\affiliation{\mbox{Institute for Quantum Materials and Technologies, Karlsruhe Institute of Technology, 76131 Karlsruhe, Germany}}
\affiliation{\mbox{Institut f\"ur Theorie der Kondensierten Materie, Karlsruhe Institute of Technology, 76131 Karlsruhe, Germany}}
\author{Alexander D.~Mirlin}
\affiliation{\mbox{Institute for Quantum Materials and Technologies, Karlsruhe Institute of Technology, 76131 Karlsruhe, Germany}}
\affiliation{\mbox{Institut f\"ur Theorie der Kondensierten Materie, Karlsruhe Institute of Technology, 76131 Karlsruhe, Germany}}

\begin{abstract}
We consider a quantum many-body lattice system that is coupled to ancillary degrees of freedom (``detectors''), which are periodically measured by means of strong projective measurements. The concentration $\rho_a$ of ancillae and their coupling $M$ to the main system are considered as parameters. We explore the dynamics of density and of entanglement entropy in the chain, for various values of $\rho_a$ and $M$ for two models of the detector-chain interaction that couple the local density in the chain to a detector degree of freedom. It is found that, for the density-density ($S_z s_z$-type in spin language) coupling, the critical values $M_c$ for the measurement-induced entanglement transition depends sensitively on $\rho_a$. 
Moreover, our results indicate that for a sufficiently small $\rho_a$ the transition in this model disappears, i.e., a finite density of detectors is needed to reach a disentangling phase. The behavior is qualitatively different for the second model, with density-hopping ($S_z s_x$-type) coupling. Specifically, the dynamics is much less sensitive to the concentration $\rho_a$ of detectors than in the first model. Furthermore, the dependence of entanglement on the coupling strength $M$ is strongly non-monotonic, indicating re-entrance of the entangling phase at large $M$.
\end{abstract}

\maketitle

\section{Introduction}

The dynamics of monitored open quantum systems \cite{Wiseman2009, Jacobs2014} has been of interest since the early days of quantum mechanics. Indeed, the Born rule providing a probabilistic interpretation of the wave function assumes that the system of interest is interacting with some external ``observer'' \cite{Zurek2003a, Schlosshauer2005a} that can induce wave-function collapse. Recently, a renewed surge of interest in the topic has emerged because of the relevance of the problem of quantum information processing \cite{Preskill2018a}. In this context, one aims at detailed understanding of the effect of external monitoring (measurements) on the otherwise unitary dynamics of a quantum many-body system.
This effect generically depends on the type and strength of interaction between the system and its environment (measurement apparatus). 

A key prediction in this field is the existence of a dynamical phase transition in monitored quantum systems.
In the absence of measurements, a generic (highly excited) state in an interacting many-body system would become highly entangled, with volume-law scaling of the entanglement entropy \cite{Page1993}. This behavior is closely related to the eigenstate thermalization hypothesis~\cite{Deutsch1991, Srednicki1994}. 
The transition is driven by the rate and strength of measurements, with sufficiently strong and frequent measurements driving the system to a disentangled state. Hence, this type of phase transitions has been dubbed a \emph{measurement-induced entanglement transition}, which has been theoretically studied in various settings \cite{Li2018a, Skinner2019a, Chan2019a, Szyniszewski2019a, Li2019a,  Bao2020a, Choi2020a, Gullans2020a, Gullans2020b, Jian2020a, Zabalo2020a, Iaconis2020a, Turkeshi2020a, Zhang2020c, Szyniszewski2020, Tang2020a, Fuji2020a, Goto2020a, Lunt2020a, Lang2020a, Rossini2020a, Cao2019a, Alberton2020a, Chen2020a, Nahum2021a, Ippoliti2021a, Ippoliti2021b, Lavasani2021a, Lavasani2021b, Sang2021a, Buchhold2021a, Biella2021a, Turkeshi2021, Tang2021a, Fisher2022,  Kelly2022a, Liu2022a, Weinstein2022a, Barratt2022, Agrawal2022, Coppola2022, Ladewig2022, Carollo2022, Buchhold2022, Yang2022, Marcin2022, Tirrito2022,  Jian2021a, Minoguchi2022, Doggen2022a, VanRegemortel2021a, Altland2022, Block2022a, Sierant2022a, Boorman2022, Minato2022a, Kalsi2022,  Sierant2023a, Doggen2023a,  Jian2023, Fava2023, Yang2023, Weinstein2023, Merritt2023, Murciano2023, Kells2023, Majidy2023, Yamamoto2023, Silveri2023, Poboiko2023a, Poboiko2023b,  Shkolnik2023, jin2023, Chahine2023, Xing2023, Lumia2023}.
Experimental studies of the measurement-induced entanglement transition have been undertaken in superconducting qubit \cite{Koh2022, Hoke2023} and trapped-ion \cite{Noel2022a, Agrawal2023} architectures.

It is understood that the volume-law behavior shows up only in the presence of interparticle interaction. At the same time, numerical modeling of interacting many-body systems is very costly from the point of view of computational resources. For this reason, most of the previous numerical works on measurement-induced transition in interacting systems were carried out for special models that are particularly convenient for exact simulations of quantum dynamics. Only a few works have addressed measurement-induced transitions in interacting models described by ``conventional'' time-independent Hamiltonians. Hence, only little is known about sensitivity of the transition in such systems to the specifics of the measurement protocol. These include implementation of measurements (e.g., projective or generalized ancilla-based, continuous or stroboscopic), type and strength of coupling between the system and ancillae, concentration of the detectors, etc.   

In a previous work \cite{Doggen2023a}, we have started to address this problem by considering an interacting chain, with every site coupled to an ancillary degree of freedom (the concentration of two-level detectors $\rho_a=1$). To overcome very stringent system-size limitations in exact simulations of quantum dynamics, we have applied the matrix-product-state (MPS) approach to the measurement-transition problem. The results exhibited clear signatures of a transition between area-law and volume-law phases, driven by strength $M$ of the system-ancilla coupling. This coupling was chosen to be of density-density type 
(i.e., of $S_z s_z$ type in the spin language), and the ancillae were projectively measured in the $s_z$ basis.
Interestingly, it was also found in Ref.~\cite{Doggen2023a} that, even if only one or two ancillae of this type are coupled to the system (close to the bipartition cut), the entanglement increase after a quench gets strongly suppressed by the measurements at a strength $M$ close to the critical $M_c$ for the transition at $\rho_a=1$. This poses a question about the dependence of $M_c$ on the concentration of detectors $\rho_a$. 

One can note a certain similarity between $\rho_a$ and the probability $p$ of measurement in models with random locations and times of measurements such as random quantum circuits where the entanglement transition driven by $p$ was originally studied  \cite{Skinner2019a, Li2018a, Chan2019a}. However, there is a clear difference between these two situations: in the first one, the positions of measurements are fixed, while in the second one, they change randomly in time. In particular, the importance of this difference manifests in the case of special one-dimensional (1D) circuit models that can be mapped to percolation~\cite{Skinner2019a, Bao2020a, Barratt2022, Agrawal2022}: while in the random case, there is a two-dimensional (2D) percolation transition, fixed positions of detectors ``cut'' the system, implying an area law for the entanglement. 

Another important question concerns the dependence of the entanglement dynamics on the measurement operator (including the ancilla Hamiltonian in the case the measurement is implemented with the help of an ancilla).
Specifically, one can choose different monitored observables in the system and also different ancilla degrees of freedom involved in the coupling to the system (for a fixed projection basis of the ancilla). It is a priori unclear how this choice would influence the measurement-induced transition.

In this paper, we investigate the sensitivity of the entanglement dynamics in an interacting chain subjected to ancilla-based measurements (Fig.~\ref{fig:diag}) with respect to the concentration $\rho_a$ of detectors and to the type of measurement.  We use a numerical approach involving the MPS formalism to study sufficiently large systems.
Starting with the model introduced in Ref.~\cite{Doggen2023a} with the density-density coupling at $\rho_a=1$ (this was the only case of finite ancilla density addressed in Ref.~\cite{Doggen2023a}) as a reference point, we first decrease $\rho_a$ in this setup to $\rho_a=1/2$ and $\rho_a=1/4$. Next, we modify the model by replacing the density-density coupling with the density-hopping one, where the occupation of the chain site affects the hopping in the ancilla. In the latter setup, we also consider ancilla concentrations $\rho_a=1, 1/2$, and $1/4$. For both models and for each value of $\rho_a$, we scan over a broad range of the coupling strength $M$. This allows us to construct qualitative phase diagrams of entanglement entropy in the parameter planes $\rho_a,\,M$ for both models.

\section{Model}
\label{sec:model}

We consider a lattice system of hard-core bosons that consists of a main chain coupled to ancillary qubits. The choice of the model is motivated by the fact that the hard-core boson chain is equivalent to a spin-1/2 chain and, hence, to a chain of qubits. Further, this model is simpler than the model of interacting spinful fermions. At the same time, it is directly applicable to experimentally studied chains of cold atoms.

\begin{figure}
\includegraphics[width=0.95\columnwidth]{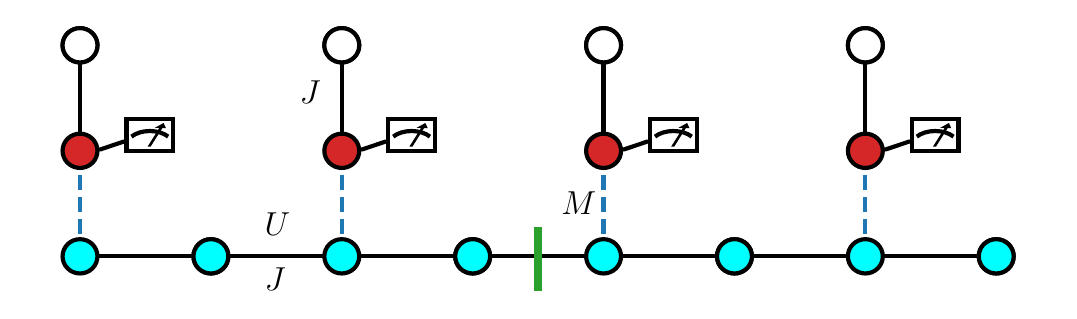} 
\vspace*{0.5cm}
\hspace*{-0.5cm}
\includegraphics[width=\columnwidth]{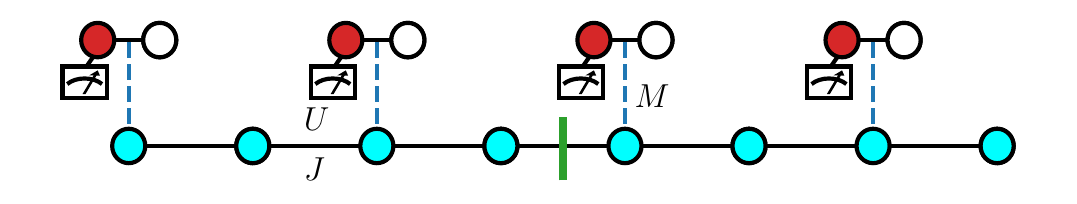}
    \caption{Schematic depiction of the setup (shown here for system size $L = 8$ and the ancilla concentration $\rho_a = 1/2$). The main chain (cyan symbols) is characterized by Hamiltonian (\ref{eq:ham}) with hopping $J$ and nearest-neighbor interaction $U$. 
    Detectors (two-level ancillas, each represented by a pair of red and blue sites) are coupled periodically to the main chain (the system-ancilla coupling indicated by green dashed lines). Projective measurements are performed on the red ancillary site at regular intervals $\Delta T$. 
    At the beginning of the protocol, the chain is initialized in the ground state of its Hamiltonian, while each ancilla is initialized in the state with its red site occupied and the blue site empty. We consider the two models of detectors: Model (i) with density-density ($S^zs^z$-type) coupling, Eqs.~\eqref{eq:meas_ham},\eqref{eq:meas_ham_sa1} (top panel); Model (ii) with density-hopping ($S^zs^x$-type) coupling, Eq.~\eqref{eq:meas_ham_sa2} (bottom panel). 
    The bipartite entropy of entanglement is computed with respect to the division in the middle of the chain (as indicated by the thick green line). In the numerical implementation of the model, the main chain and
ancillary pairs are mapped onto a single 1D chain, with the ancilla sites folded to the right of the corresponding monitored site. The sequence of colors in the resulting chain for the example with $\rho_a=1/2$ is as follows (from left to right): cyan, red, white, cyan, cyan, red, white, and so on.}
    \label{fig:diag}
\end{figure}

The main chain is defined by the Hamiltonian
\begin{equation}
\mathcal{H}_\mathrm{s} =  \sum_{i=1}^{L-1} \left[ -\frac{J}{2} \left(b_{i}^\dagger b_{i+1} + \mathrm{H.c.}\right)  + U\hat{n}_{i} \hat{n}_{i+1} \right],
\label{eq:ham}
\end{equation}
where $L$ is the chain length, $b^{\dagger}_i$ creates a hard-core boson on lattice site $i$, $\hat{n}_i \equiv b_i^\dagger b_i$ is the density, 
$J$ is a hopping parameter, and $U$ is the nearest-neighbor interaction strength. Below we set $J=1$ to fix the energy units.
The measurement procedure is implemented by performing projective density measurements on the ancillary sites coupled to the main system, as schematically depicted in Fig.~\ref{fig:diag}. 
We consider two different types of coupling of the ancillae to the main system. 

The first measurement protocol [model (i)] involves a density-density interaction between the system and the ancillae, with the total Hamiltonian given by
\begin{equation}
\mathcal{H} = \mathcal{H}_\mathrm{s} + {\sum_j}^\prime\left[\mathcal{H}_\mathrm{a}^{(j)} + \mathcal{H}_\mathrm{sa}^{(j)}\right], 
\label{eq:H-total}
\end{equation}
where the sum goes over sites coupled to ancillae, where
\begin{equation}
\mathcal{H}_\mathrm{a}^{(j)} =  -\frac{1}{2}\left(a_{j,1}^\dagger a_{j,2} + a_{j,2}^\dagger a_{j,1} \right) \quad \mathrm{model \, (i)}
\label{eq:meas_ham}
\end{equation}
is the own ancilla Hamiltonian (with the hopping parameter equal to that in the main chain), and 
\begin{equation}
\mathcal{H}_\mathrm{sa}^{(j)} = - M \hat{n}_{j} a_{j,1}^\dagger a_{j,1} \quad \mathrm{model \, (i)}
\label{eq:meas_ham_sa1}
\end{equation}
describes the system-ancilla coupling.
Here  $a_{j,1}, a_{j,2}$ are annihilation operators of a hard-core boson in the ancillary pair.

For the second measurement protocol [model (ii)], the Hamiltonian has the same form \eqref{eq:H-total}, where now $\mathcal{H}_\mathrm{a}^{(j)}=0$ and $\mathcal{H}_\mathrm{sa}^{(j)}$ describes a 
density-hopping coupling:
\begin{equation}
    \mathcal{H}_\mathrm{sa}^{(j)} = - \frac{M}{2} \hat{n}_{j} \left( a_{j,1}^\dagger a_{j,2} + a_{j,2}^\dagger a_{j,1} \right), \quad \mathrm{model \, (ii)}.
    \label{eq:meas_ham_sa2}
\end{equation}
In this setup, the hopping between the ancilla sites is mediated by the coupling to the density on the main chain: the ancilla sites are connected only when the corresponding system site is occupied. This also means that the ancillae are effectively frozen in the limit of zero coupling $M \rightarrow 0$. 

For both models, a projective Born-rule measurement of the density $a_{j,1}^\dagger a_{j,1}$ on ancilla sites 1 is performed at regular intervals $\Delta T$. When the model \eqref{eq:ham} is represented in terms of spin-$1/2$ operators $\mathbf{S}_i$, one can view the coupling \eqref{eq:meas_ham_sa1}
 in setup (i) as $S^z s^z$ coupling, and the coupling
 \eqref{eq:meas_ham_sa2}
 in setup (ii) as $S^zs^x$ coupling, where $\mathbf{s}$ refers to the ancilla and $s_x$ corresponds to hopping between the ancilla sites.

We consider the ancilla concentrations $\rho_a=1$, 1/2, and 1/4, where $\rho_a$ is equal to the number of ancilla pairs divided by the length $L$ of the main chain. The number of ancillae therefore scales extensively with $L$, and the total number of lattice sites is \mbox{$L_\mathrm{tot} = (1 + 2\rho_a)L$}. We choose $L$ divisible by 4 and attach the ancillae starting from the first site, see Fig.~\ref{fig:diag}.

The final ingredient of the protocol is absence of resetting of the ancillae after measurements \footnote{Measurement model (ii) for $L=2$ (single-particle main chain) was considered in Refs.~\cite{Doggen2022a} and \cite{Poepperl2023} in the presence of resetting}. This induces feedback loops, with the measurement of the ancilla affecting the dynamics in the main system, which in turn affects consecutive measurement results, and so on. When the measurement frequency and the ancilla Rabi oscillation frequency are commensurate, this leads to very long-lived correlations between measurements, that can significantly affect the dynamics of the system. In Ref.~\cite{Doggen2023a}, this was observed for setups with one and two detectors (with density-density coupling) and dubbed the \emph{quantum-Zeno-valve effect} (QZVE). For $\rho_a=1$, this effect was smeared out.
Here we will investigate whether this effect shows up in the entanglement dynamics for smaller $\rho_a$ and also for an alternative [model (ii)] coupling~\footnote{Upon inspection of Eqs.~\eqref{eq:meas_ham_sa1}-\eqref{eq:meas_ham_sa2}, one can observe that in the mean-field limit $\hat{n} \rightarrow \langle \hat{n} \rangle$ there is a well-defined Rabi oscillation frequency for the ancilla boson in both setups (i) and (ii).}.

The dynamics is computed using the time-dependent variational principle \cite{Haegeman2016a}, with a procedure similar to the one used in Ref.~\cite{Doggen2023a}. The whole setup (main chain and ancillary pairs) is mapped onto a 1D chain using matrix product operators, where the ancillae are ``folded'' into the chain on consecutive sites, and the main chain sites are then coupled through next-next-nearest neighbor terms. The numerical approach is based on MPS \cite{Schollwock2011a, Paeckel2019a}, a type of tensor network wherein we consider a variational subspace of the whole Hilbert space of the system. The particle-number conservation is automatically respected within this approach; we further constrain the Hilbert space to the subspace with a fixed total particle number \cite{tenpy}, which greatly accelerates computations. The total Hilbert space of the model scales as $\propto 2^{L_\mathrm{tot}}$; in the MPS approximation, we consider only a subspace with polynomial complexity, controlled by a numerical parameter $\chi$ called the bond dimension. The lower computational complexity of the cases $\rho_a=1/2$ and 1/4 allows us to consider a larger size of the main chain ($L=40$) than in Ref.~\cite{Doggen2023a}.

\section{Results}

\subsection{Observables}

We focus on the following observables. To quantify entanglement, we compute the von Neumann bipartite entropy of entanglement $S(t)$, where the bipartition is taken in the middle of the main system as indicated in Fig.~\ref{fig:diag},
\begin{equation}
S(t)  = -\mathrm{Tr}\Big( \rho_A \ln \rho_A \Big), \quad \rho_A = \mathrm{Tr}_\mathrm{B} \rho.
\end{equation}
Here $\rho$ is the density matrix of the whole system (including the detectors), and $\rho_A$ is the reduced density matrix corresponding to a part $A$ of a bipartition in parts $A$ and $B$.
Furthermore, we track the dynamics of the density $n_i(t) \equiv \langle \hat{n}_i \rangle (t)$ at every site in the system (where $\langle \ldots \rangle$ denotes the averaging over a quantum state), as a function of time $t$. From this, we can also compute the probability density function $P(n;t)$, which quantifies the fluctuations of the density throughout the system.

For each choice of parameters $\rho_a$ and M, we numerically compute an ensemble (typically $\sim 40$, in some cases up to 200) quantum trajectories
with the measurement interval $\Delta T=2$. In what follows, we use the MPS bond dimension $\chi=128$, which establishes an upper cutoff for the entanglement entropy $S=\ln 128 \simeq 4.85$.

\begin{figure*}
    \centering
    \includegraphics[width=0.9\columnwidth]{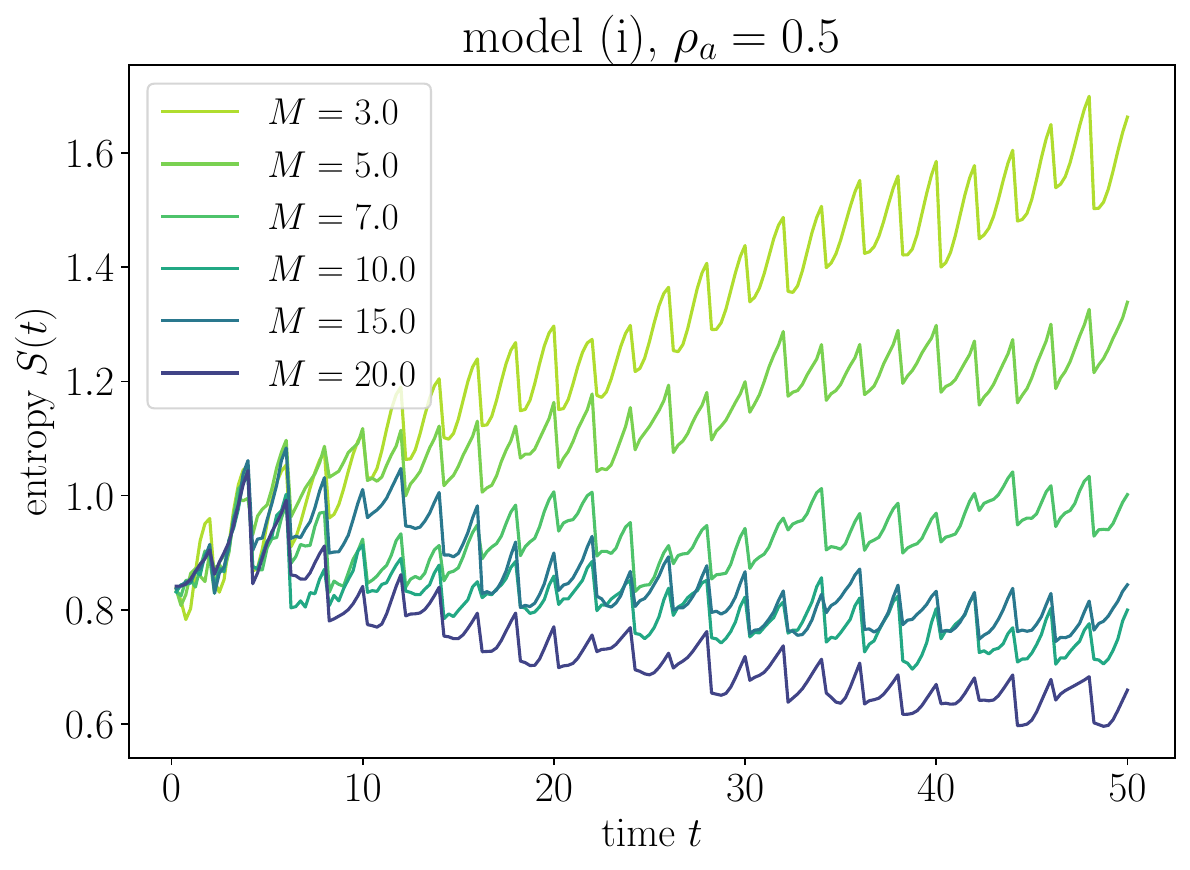}
    \includegraphics[width=0.9\columnwidth]{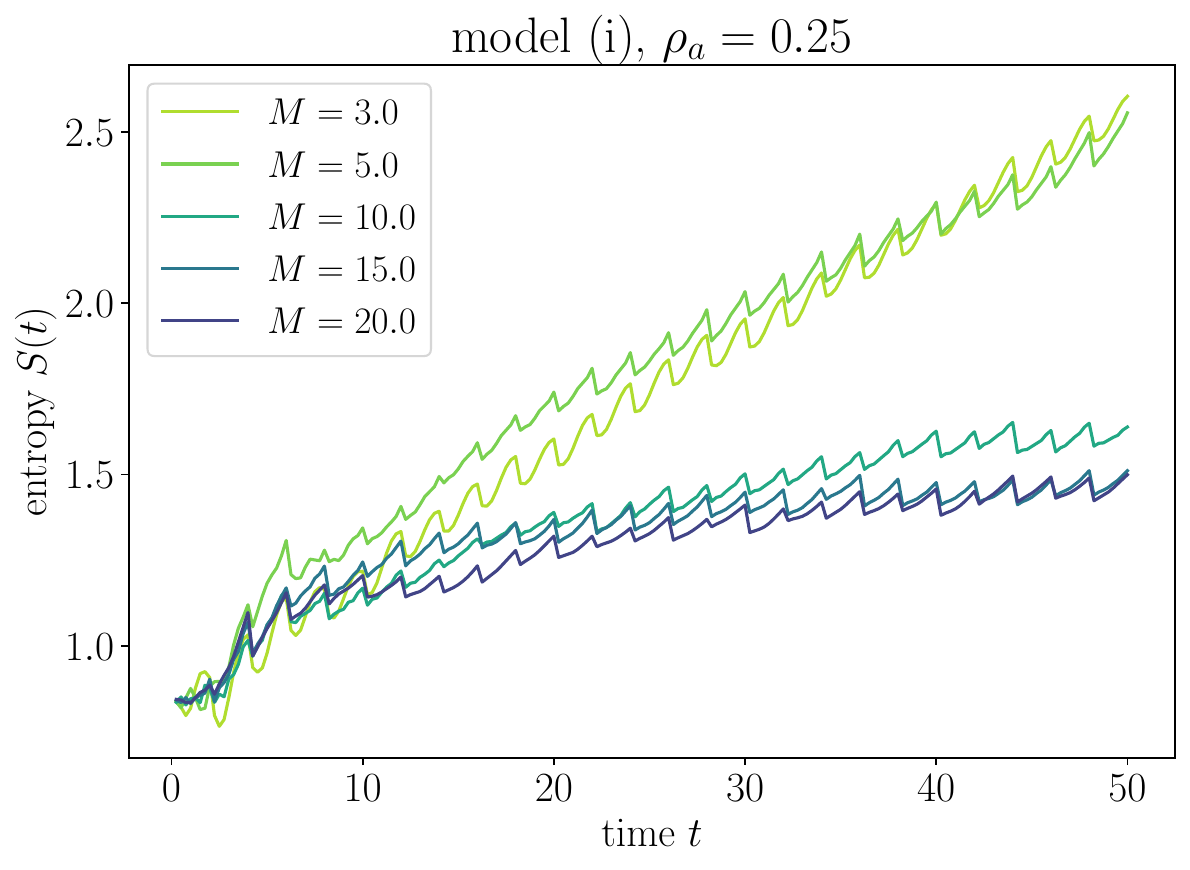}
    \caption{Average entanglement entropy $S(t)$ for model (i) at various values of the system-ancilla density-density coupling $M$ for system length $L = 40$. Left panel: $\rho_a = 1/2$; an entangling trend is observed for coupling strength $M=3$ and 5, while disentangling behavior is observed for stronger coupling, $M=10$, 15, and 20; {\color{black} for $M=7$, the entanglement entropy stays approximately constant as a function of time.} Right panel: $\rho_a=1/4$; all curves show an entangling trend.}
    \label{fig:ent_model1}
\end{figure*}

\subsection{Setup (i): Density-density coupling}

We first consider model (i) with the density-density interaction, as defined in Sec.~\ref{sec:model}.
We start with the case where there is an ancilla pair at every site, $\rho_a=1$, as in Ref.~\cite{Doggen2023a}. 
(Note that our definition of the system-ancilla coupling is different by the sign from that in  Ref.~\cite{Doggen2023a}. This, however, does not lead to any essential difference in the results.)
For $\rho_a=1$, the data provide evidence for a measurement-induced entanglement transition at $M_c\approx 5$, see Ref.~\cite{Doggen2023a}.
Specifically, $M > M_c$ leads to disentangling behavior, with $S(t)$ saturating at a relatively small value, smaller than $S(t=0)$ and way smaller than the cutoff set by the MPS bond dimension $\chi$. On the other hand, for $M < M_c$, the entanglement entropy $S(t)$ grows with time, providing an indication of the volume law. For sufficiently large systems and moderate times, the volume-law phase manifests itself in the linear growth of $S(t)$ with time until it approaches the saturation cutoff set by the bond dimension $\chi$.

We now proceed by analyzing the effect of reduction of the ancilla concentration $\rho_a$. In Fig.~\ref{fig:ent_model1}, we show the entanglement entropy $S(t)$ averaged over quantum trajectories for $\rho_a=1/2$ and $\rho_a=1/4$. First of all, the average entropy is higher for smaller $\rho_a$, so that having fewer ancillae disentangles less, as may be expected. From this, we can infer that the transition, if any, will be at a larger value of $M$.

For $\rho_a=1/2$, the behavior of entanglement is qualitatively similar to that for $\rho_a=1$: the entangling behavior for small $M$ and disentangling behavior at large $M$ are clearly visible. 
This provides an indication of the entanglement transition at $M_c\approx 7-8$ (see also the phase diagram in Sec.~\ref{sec:phase} below and plots in Supplemental Material \cite{SupMat}). 
Importantly, this value of $M_c$ is significantly larger than the critical value $M_c\approx 5$ observed for $\rho_a=1$. Thus, $M_c$ in model (i) strongly depends on the ancilla concentration $\rho_a$. 

To further explore the evolution of entanglement with decreasing $\rho_a$, we show in the right panel of Fig.~\ref{fig:ent_model1} $S(t)$ for $\rho_a=1/4$. We observe a drastic change in its behavior: all curves show that $S(t)$ increases with time, even for such a large coupling as $M=20$. Although for large values of $M$ the growth is relatively slow (and we cannot exclude saturation at longer times), we do not have clear evidence of the transition at all. This further confirms that $M_c$ increases with decreasing $\rho_a$ and probably diverges at some $\rho_a$ not far from 1/4. 

In both cases of $\rho_a=1/2$ and $\rho_a=1/4$, the entropy in Fig.~\ref{fig:ent_model1} is well below the maximum 
entanglement reachable by the MPS, $S\approx 4.85$. 
Importantly, the most entangling curves show a nearly linear growth characteristic of the volume-law phase, without any signature of bending down, up to the maximum value at $t=50$. This observation allows us to conclude that the chosen finite bond dimension is sufficient for studying entanglement for the parameters of Fig.~\ref{fig:ent_model1}. As we show in Sec.~\ref{sec:setup2} below {\color{black}and in Appendix \ref{appendix}}, the MPS cutoff is largely inessential for the values of the entanglement entropy below $S\approx 3$.

\begin{figure*}
    \centering
    \includegraphics[width=\textwidth]{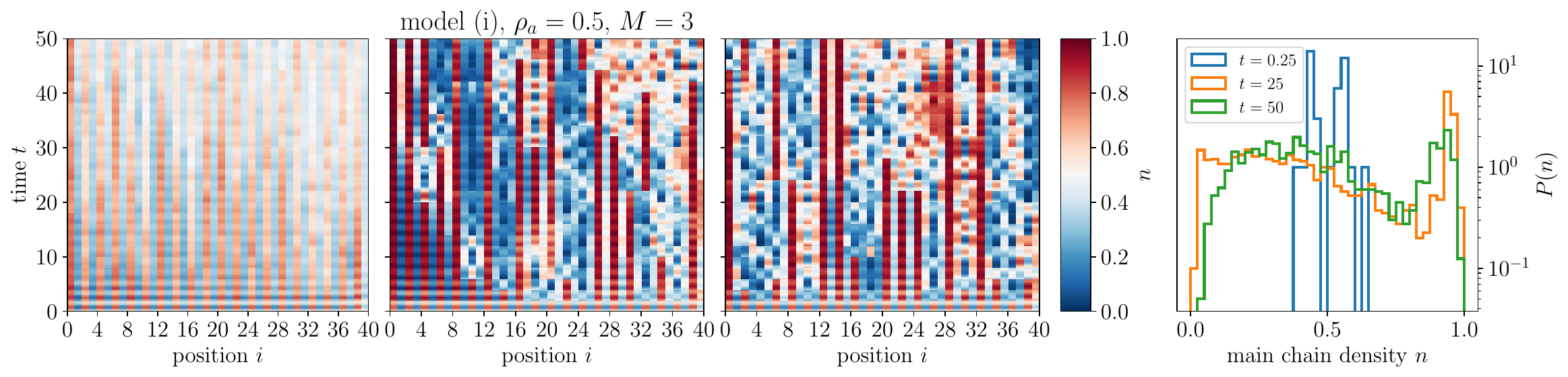}
    \caption{Time evolution of the density $n_i(t)$ for model (i) with parameters $\rho_a = 1/2,\ M = 3,\ L = 40,\ \Delta T = 2$. 
    The coupling strength here is relatively weak, corresponding to the most entangling entropy curve in the left panel of Fig.~\ref{fig:ent_model1}.
    First panel: $n_i(t)$ averaged over the ensemble of quantum trajectories; second panel: $n_i(t)$ for the quantum trajectory with the smallest $S(t=50)$; 
    third panel: $n_i(t)$ for the quantum trajectory with the largest $S(t=50)$; fourth panel: density distribution function $P(n;t)$ for $t=0.25,\ 25$, and $50$ shows broadening with time, as well as the appearance of a peak near $n=1$ (red stripes in the other panels).}
    \label{fig:M3_rhoa05}
\end{figure*}

\begin{figure*}
    \centering
    \includegraphics[width=\textwidth]{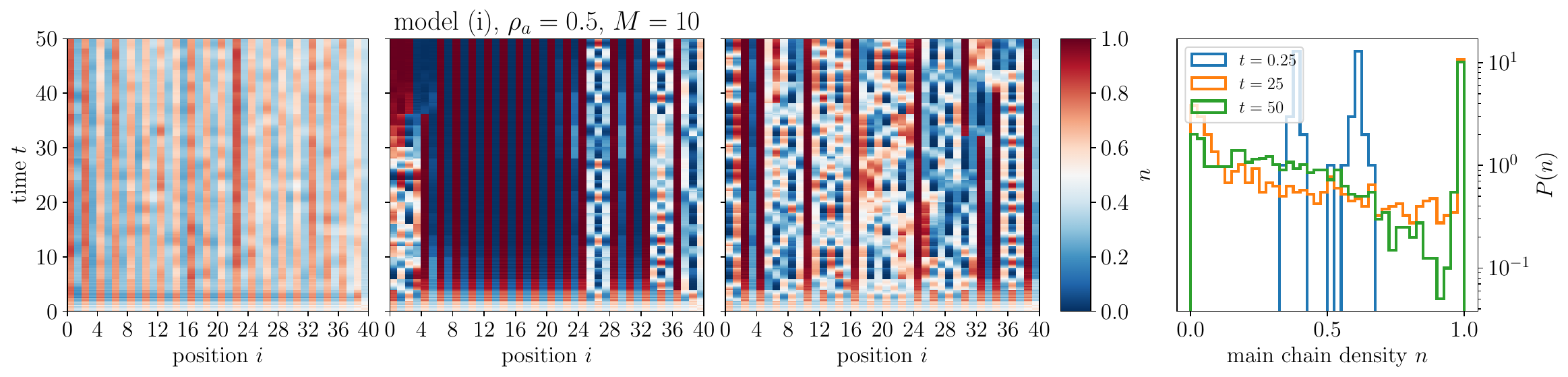}
    \caption{Evolution in time of the density for model (i) with $\rho_a = 1/2,\ L = 40,\ \Delta T = 2$, 
    as in Fig.~\ref{fig:M3_rhoa05},
    but for a stronger ancilla-chain coupling $M = 10$, corresponding to a disentangling entropy curve in the left panel of Fig.~\ref{fig:ent_model1}.
    Compared to the case of weaker coupling,  Fig.~\ref{fig:M3_rhoa05}, the density pattern for the least entangled quantum trajectory (second panel) reveals extensive frozen regions (alternating dark red and blue stripes forming a ``pajama'' structure); the pattern of the most entangled trajectory (third panel) has a higher level of contrast than for $M=3$. These features are reflected in the appearance of two peaks, near $n=0$ and $n=1$ in the density distribution function $P(n)$ (fourth panel), as well as by the absence of a maximum around $n=0.5$ at longer times.}
    \label{fig:M10_rhoa05}
\end{figure*}
\begin{figure*}
    \centering
    \includegraphics[width=\textwidth]{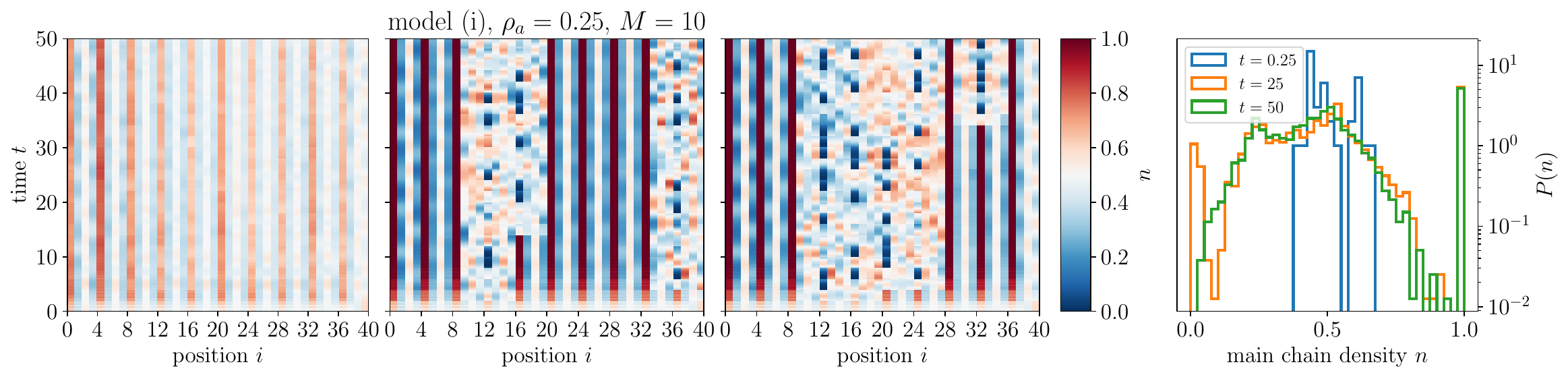}
    \caption{Evolution of the density for model (i) with $M = 10,\ L = 40,\ \Delta T = 2$, as in Fig.~\ref{fig:M10_rhoa05}, but for a lower concentration of ancilla pairs, $\rho_a = 1/4$, corresponding to a disentangling entopy curve (right panel of Fig.~\ref{fig:ent_model1}). Lowering the ancilla concentration reduces the contrast of the individual density patterns, which is reflected by the bell-shape distribution $P(n)$ in the fourth panel, with the maximum around $n=0.5$; at longest times a narrow peak in $P(n)$ near $n=1$ is developed. The difference between the patterns for the least (second panel) and most (third pattern) entangling trajectories is much less pronounced for $\rho_a=1/4$ than for $\rho_a=1/2$.}
      \label{fig:M10_rhoa025}
\end{figure*}

In order to shed more light on the physics behind the entanglement dynamics, we illustrate in Figs.~\ref{fig:M3_rhoa05}-\ref{fig:M10_rhoa025} the temporary evolution of the density $n_i(t)$ for $\rho_a=1/2$ at relatively weak coupling $M=3$ and relatively strong coupling $M=10$, as well as for $\rho_a=1/4$ at $M=10$ (plots for other representative values of the parameters can be found in Supplemental Material \cite{SupMat}).
In each of the figures, the left panel shows $n_i(t)$ averaged over the ensemble of quantum trajectories, while the second (third) panel shows $n_i(t)$ for the quantum trajectory with the smallest (respectively, largest) value of $S(t=50)$. The two extreme cases of quantum trajectories differ in the contrast of density patterns. Importantly, for the most disentangling quantum trajectories, the bipartition cut for the entanglement entropy is located within the spatial region characterized by the high contrast in the density pattern. This relates the entropy with the density fluctuations, see below. Finally, the fourth panel presents the density distribution function $P(n)$ at times corresponding to the beginning, midpoint, and end of evolution.

Compared to the corresponding plots for $\rho_a=1$ (see Ref.~\cite{Doggen2023a}), we see a striking difference. While for $\rho_a=1$ clusterization of the density was found at large $M$, here we observe a well-pronounced striped structure. This is related to the fact that the ancillae are coupled only to a subset of sites of the main chain. Long-lived metastable ``pajama'' structures consisting of vertical stripes of alternating red ($n\approx 1$) and blue ($n\approx 0$) colors form beyond the size of a single site. (The clearest example of the striped  ``pajama'' density modulations is seen in the second panel of Fig.~\ref{fig:M10_rhoa05} between sites 6 and 22.) These structures prevent entanglement growth and can be attributed to the QZVE, discovered in Ref.~\cite{Doggen2023a} for setups with one or two ancillae. 

The QZVE originates from the formation of a quasi-bound state of particles in the sites coupled by the system-ancilla interaction ~\cite{Doggen2023a}. The bound state immobilizes the particle in the main chain, preventing other particles from crossing the occupied site. The site of the main chain involved in the bound state is then occupied, while the neighboring sites are largely empty, giving rise to the red-blue pajama-like stripes in Figs.~\ref{fig:M3_rhoa05}-\ref{fig:M10_rhoa025}. As a result, both the density and entanglement dynamics become blocked by the bound state. This blocking is particularly important when the frozen region overlaps with the bipartition cut used to calculate the entanglement entropy, as exemplified in the second panels of Figs.~\ref{fig:M3_rhoa05}-\ref{fig:M10_rhoa025}: the emergent barrier effectively cuts the main chain into two.

Further measurements of the ancilla site typically find the ancilla particle bound in this state, which maintains this configuration {\color{black}(hence, ``quantum Zeno")}. Very rarely, the measurements of the ancilla pair may nevertheless find the red site (see Fig.~\ref{fig:diag}) empty, which immediately breaks the quasi-bound state and removes the barrier for dynamics in the main chain (hence, ``valve"). Similarly, the blocking regions can be established by the measurements, when the particle is measured in the ancilla site
while the corresponding site of the main chain is occupied. This is the origin of metastability of some pajama stripes seen in the panels showing individual trajectories in Figs.~\ref{fig:M3_rhoa05}-\ref{fig:M10_rhoa025} (see, e.g., the red stripe in the second panel of Fig.~\ref{fig:M10_rhoa05} at site $i=3$, which emerges at $t=36$, and the red stripe in the third panel of Fig.~\ref{fig:M10_rhoa025} near site $i=32$, which abruptly terminates at $t=34$). In Supplemental Material \cite{SupMat}, we show, in addition to the density patterns in the main chain, the evolution of occupation of the measured ancillary sites. The QZVE is clearly seen there in the correlations between the long-living stripes in the main chain and ancillae.

\begin{figure*}[t!]
    \centering
    \includegraphics[width=.67\columnwidth]{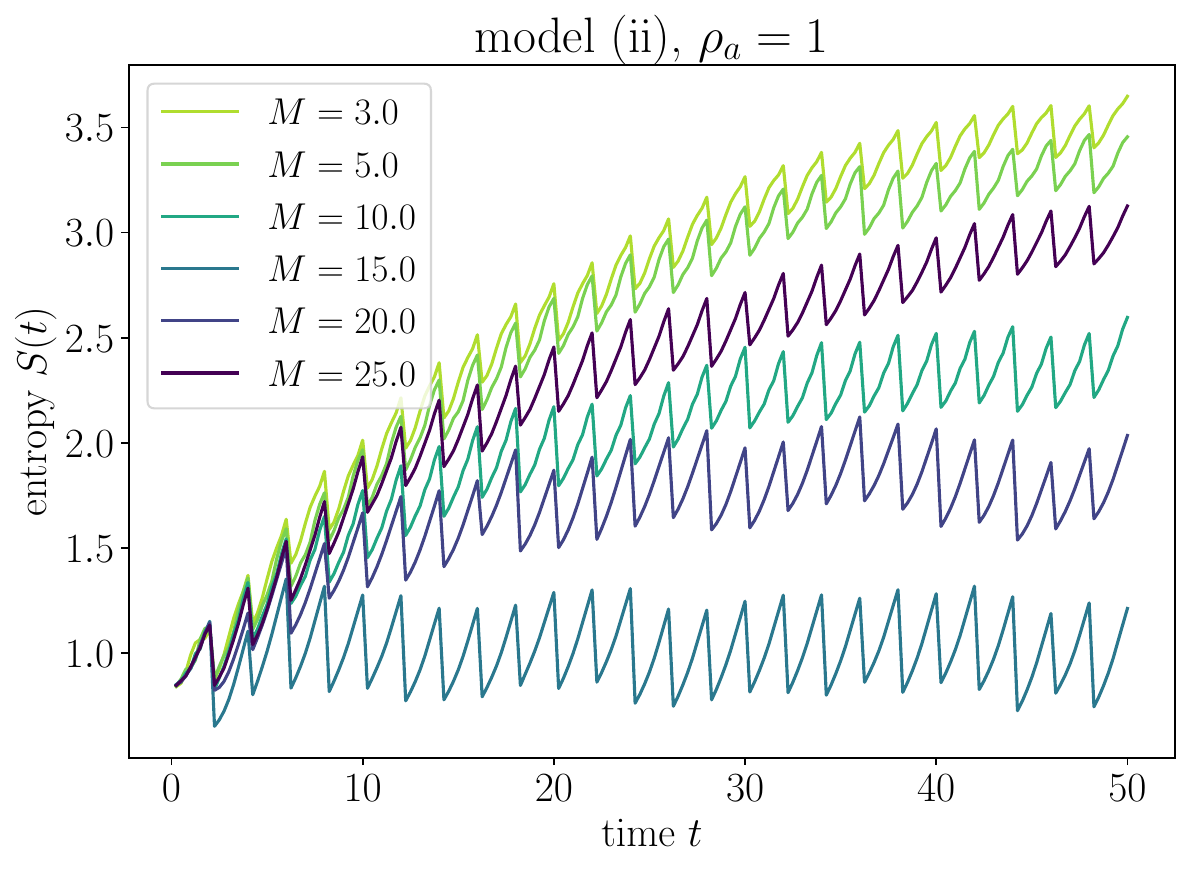}
    \includegraphics[width=.67\columnwidth]{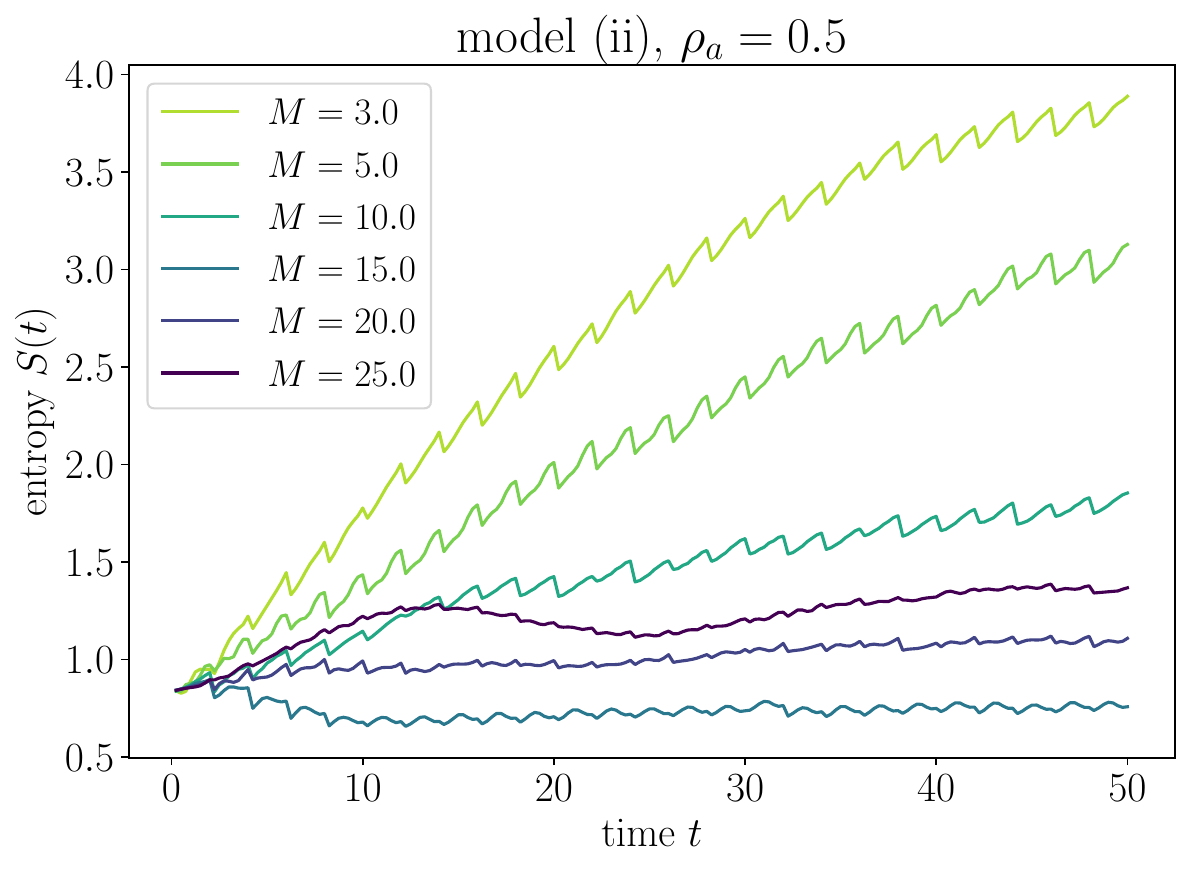}
    \includegraphics[width=.67\columnwidth]{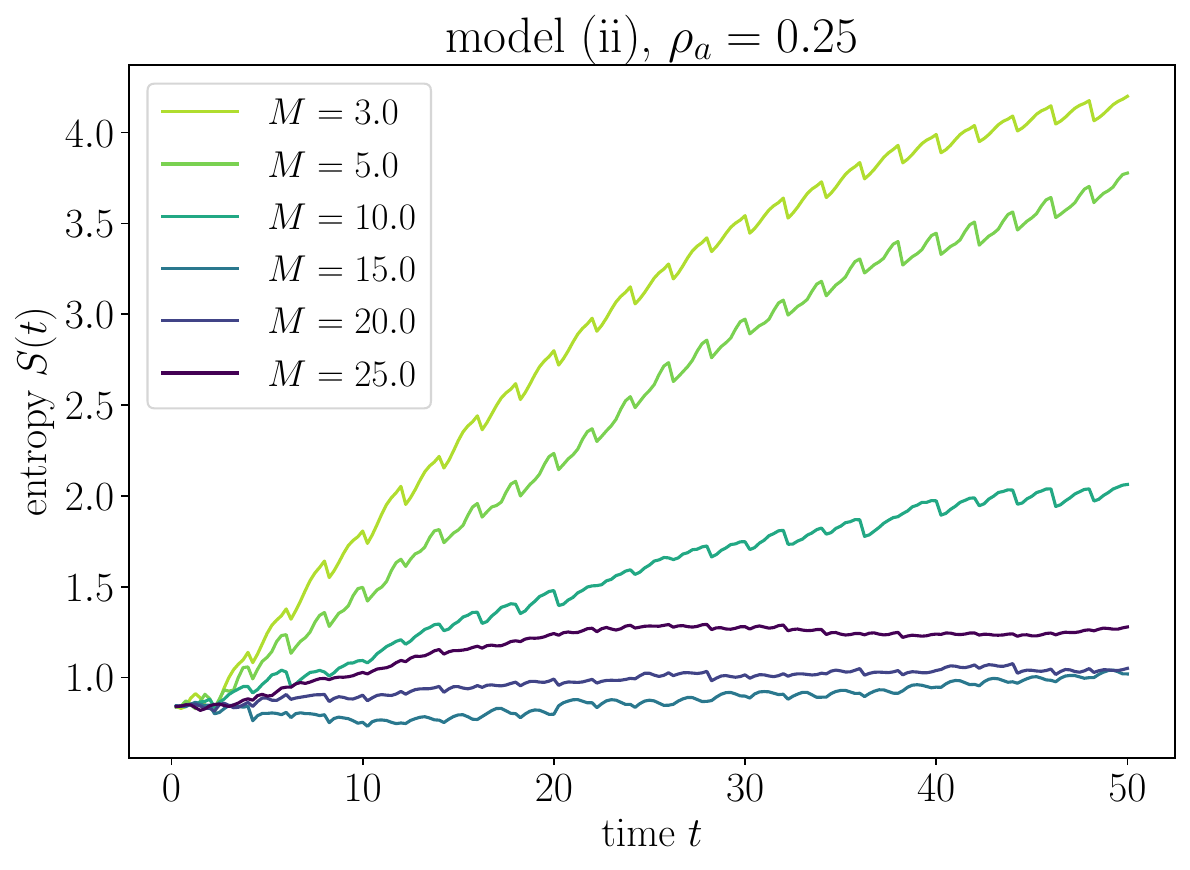}
    \caption{Average entanglement entropy $S(t)$ for model (ii) at various values of the system-ancilla density-hopping coupling $M$ for $L = 40$. Left panel: $\rho_a = 1$; middle panel: $\rho_a = 1/2$; right panel: $\rho_a=1/4$. All three panels demonstrate a non-monotonic dependence of $S(t=50)$ on the coupling strength, with the most disentangled curves corresponding to $M=15$. }
    \label{fig:ent_model2}
\end{figure*}

For both cases $\rho_a=1/2$ and $\rho_a=1/4$, the contrast of plots for individual trajectories is enhanced with increasing $M$, which implies stabilization of the QZVE. 
We further note that, for not too strong coupling $M=3$, the least and most entangled quantum trajectories produce similar density plots.
On the contrary, for $M=10$, we observe much stronger fluctuations within the ensemble of quantum trajectories, as visualized by the striking difference between the second and third panels of Fig.~\ref{fig:M10_rhoa05}.

For $\rho_a=1/4$, the effect of measurements on the density profiles is considerably weaker: even for $M=10$, the least and most entangled trajectories are characterized by visually similar densities, see the second and third panels in Fig.~\ref{fig:M10_rhoa025}. The large difference in entanglement in these two trajectories is because the bipartition cut in the third panel (most entangled trajectory) occurs in the middle of the pajama-free region.

The averaged (over the ensemble of quantum trajectories) densities shown in the left panels of Figs.~\ref{fig:M3_rhoa05}-\ref{fig:M10_rhoa025} also exhibit pajama-like patterns but with much weaker contrast. This reduction of contrast originates from the two possibilities of the metastable frozen states formed by the ancilla site and the site of the chain to which it is attached: either an ancilla-particle or an ancilla-hole quasi-bound state. This is well seen in all panels showing individual quantum trajectories, where the red vertical stripes correspond to particles and the blue stripes to holes. Red stripes are somewhat more stable, leading to residual pajamas in the panels for averaged densities.

Averaging of the density profiles (or other ``conventional'' observables) over quantum trajectories (i.e., over sequences of measurement outcomes) can be  expressed in terms of the averaged density matrix whose evolution is described by the Lindblad equation. In the field-theoretical representation of the measurement problem, such averaged quantities correspond to replica-symmetric correlation functions, whereas the information about the measurement-induced transitions is contained in replica-asymmetric correlations that are nonlinear in the density matrix~\cite{Poboiko2023a}. The difference observed when comparing the first panels of Figs.~\ref{fig:M3_rhoa05}-\ref{fig:M10_rhoa025} with the second and third panels is exactly of this origin.

The qualitatively different physics of quantum states at small and large $M$ is reflected also in the distribution function $P(n;t)$ of $n(t)$, see the fourth panels in Figs.~\ref{fig:M3_rhoa05}-\ref{fig:M10_rhoa025} and plots in Supplemental Material~\cite{SupMat}. For sufficiently small $M$, this distribution is relatively narrow and peaked at $n=1/2$. On the other hand, at large $M$ the distribution broadens over the whole range $[0,1]$ of densities, with peaks emerging at $n=0$ and $n=1$ that reflect the high-contrast pajama structure.
The emergence of these peaks in $P(n;t)$ is correlated with the suppression of the entanglement growth, see the above discussion of the QZVE.

Finally, we recall that the initial state in our protocol is chosen so that the measured ancilla site is occupied at $t=0$, i.e., $s^z = +1$. 
If the ancillae were prepared in a different initial state, it would not affect qualitatively the long-time dynamics in model (i).
Indeed, the measurements of the ancillae are always performed in the $s^z$ direction, without resetting the ancilla pair after the measurement. Since the quasi-bound state responsible for the QZVE is formed only for the $s^z = +1$ state of the ancilla pair, the initial state of the ancilla is not essential (does not affect the state of the system after a long time). 
If the ancilla pair is initialized in the $s^z = -1$ (empty measured ancilla site), the hopping between the ancilla sites is not blocked and the ancilla will be eventually measured in the $s^z = +1$ state (our initial state), which will be then maintained by the QZVE. Similar considerations apply to any other initial states and result only in a small delay of the setting-in of the dynamics described in this section.

\begin{figure*}
    \centering
    \includegraphics[width=\textwidth]{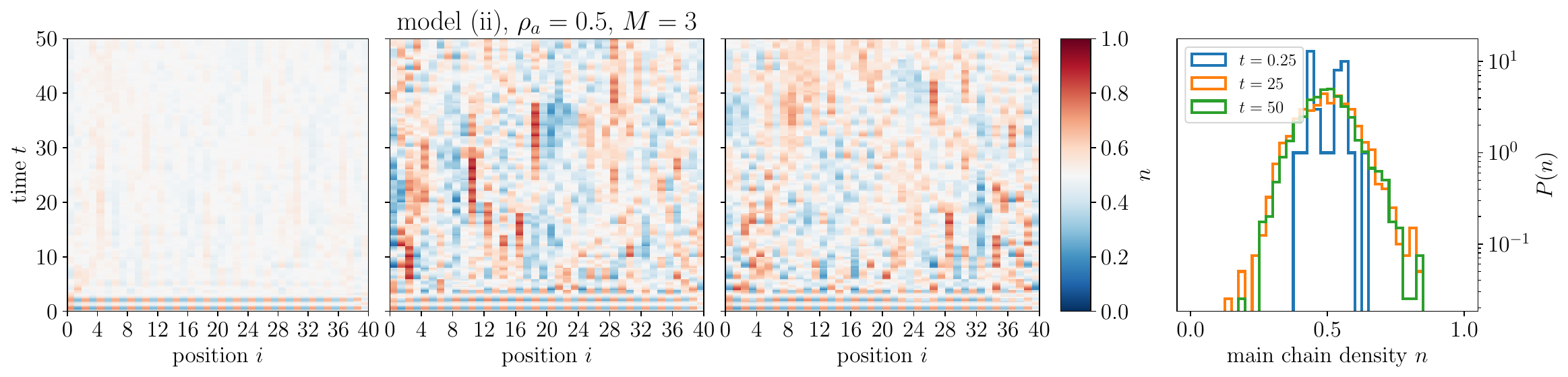}
    \caption{Evolution of density for the same parameters as in Fig.~\ref{fig:M3_rhoa05}, $\rho_a = 1/2, M = 3, L = 40$, and $\Delta T = 2$, but now for model (ii).
    The coupling strength corresponds to the most entangling entropy curve in the middle panel of Fig.~\ref{fig:ent_model2}.
    The contrast of the density patterns is significantly reduced compared to that for model (i) in Fig.~\ref{fig:M3_rhoa05}.   
    The patterns for the least and most entangling quantum trajectories (second and third panels, respectively) are now qualitatively similar. The density distribution $P(n)$ for longer times has a pronounced bell-like shape that does not change with time, with no peaks at $n=0$ and $n=1$.}
    \label{fig:M3_rhoa05_model2}
\end{figure*}
\begin{figure*}
    \centering
    \includegraphics[width=\textwidth]{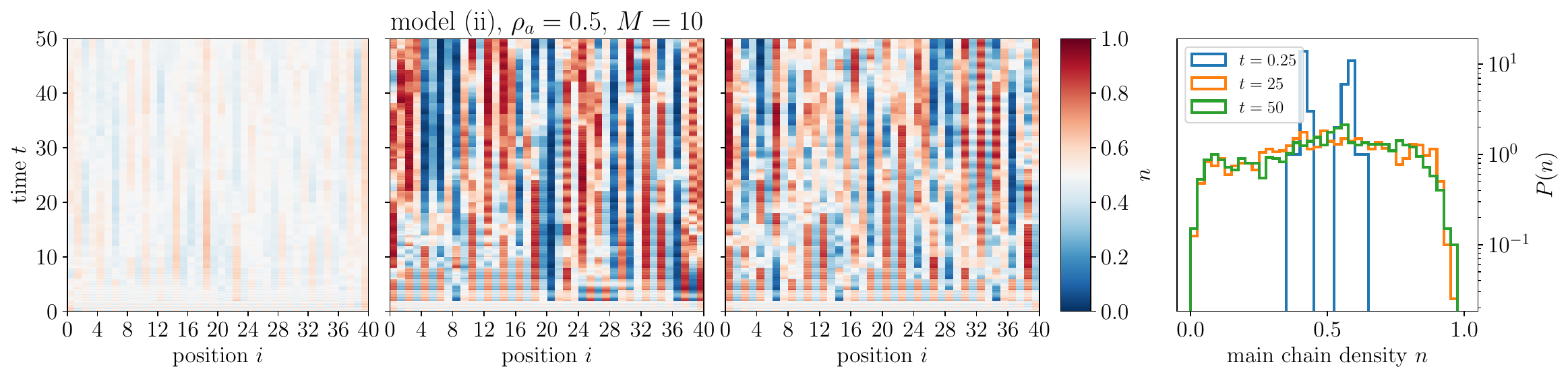}
    \caption{Evolution of density  for model (ii) with $\rho_a = 1/2,\ L = 40,\ \Delta T = 2$, as in Fig.~\ref{fig:M3_rhoa05_model2}, but for a stronger coupling, $M=10$. The contrast of the patterns is increased, as reflected by the flattening of the distribution $P(n)$ with time in the fourth panel. Still, no peaks are developed in $P(n)$ at $n=0$ and $n=1$, as opposed to Fig.~\ref{fig:M10_rhoa05} for model (i) with the same parameters. Contrary to model (i), this set of parameters for model (ii) yields entangling behavior of entropy, see Fig.~\ref{fig:ent_model2}. } 
    \label{fig:M10_rhoa05_model2}
\end{figure*}

\subsection{Setup (ii): Density-hopping coupling}
\label{sec:setup2}

We turn now to model (ii), in which the density on a site of the main chain is coupled to ancilla hopping, see Sec.~\ref{sec:model}. The time dependence of the average entanglement entropy for this model with $\rho_a=1/2$ and 1/4 and various values of coupling $M$ is shown in
Fig.~\ref{fig:ent_model2}.
For the most entangling curves, we observe a nearly linear growth up to $S\approx 3$, after which the curves start bending down as a result of proximity to the maximum possible value $S\approx 4.85$ imposed by the chosen bond dimension $\chi=128$ of the MPS approach. Thus, the entropy curves below $S\approx 3$ are largely unaffected by the MPS cutoff, similar to the curves in Fig.~\ref{fig:ent_model1}. {\color{black} This conclusion is confirmed in Appendix~\ref{appendix} by comparing the curves obtained with $\chi=128$ and higher bond dimension $\chi=256$.}

Comparing to Fig.~\ref{fig:ent_model1}, we observe an essential difference between the models (i) and (ii). While an initial increase of $M$ in model (ii) leads to a suppression of the entanglement, a re-entrant behavior is observed when $M$ grows further. This non-monotonic behavior can be related to the ``frustration'' between the coupling term and the measurement in model (ii): in the spin language, the former involves the $s^x$ component and the latter the $s^z$ component of ancilla spin.  While the measurements try to freeze the ancilla in the $z$ basis, the large-$M$ coupling leads to rapid oscillations in this basis, precluding the formation of a quasi-bound state, in contrast to model (i). It is only for certain special (``commensurability'') conditions on the product $M\Delta T$ that resonant dynamics may take place: when the hopping within the ancilla pair yields the same configuration of the ancilla spin after the measurement time interval $\Delta T$, the dynamics appears ``stroboscopically frozen''. This is similar to the commensurability effects studied in Ref.~\cite{Poepperl2023} in a ``toy model" of a monitored qubit. In the present model of a large correlated system, such commensurability effects are washed out by the correlated dynamics within the main chain: different ancilla pairs are not independent of each other. Therefore, for much larger values of $M$ no further reentrant behavior is expected, in contrast to the case of a single qubit.  
Another clear difference is that the results in model (ii) are much less sensitive to the ancilla concentration.

Results for the density evolution in model (ii) are presented in Figs.~\ref{fig:M3_rhoa05_model2} and \ref{fig:M10_rhoa05_model2} for $\rho_a=1/2$. 
For relatively small coupling, $M=3$, we observe a very low contrast: the density is typically close to its average value $n_i=1/2$, as also seen in the distribution function $P(n;t)$, see also Supplemental Material~\cite{SupMat}. This is in correspondence with a fast increase of the entropy $S(t)$ in the left panel of Fig.~\ref{fig:ent_model2}. On the other hand,  for $M=10$ a pajama-like pattern is seen as for the model (i). There is, however, a difference as compared to model (i): the long-lived metastable structures in the case of model (ii) are statistically particle-hole symmetric (red and blue stripes in the color density plot have similar appearance). In terms of the $P(n;t)$, this manifests itself in the approximate symmetry $n \to 1-n$. As in model (i), the initialization of the ancilla pair in a state different from $s^z=+1$ is not expected to change the dynamics, since the $S^z s^x$ system-ancilla coupling does not discriminate between $s^z=\pm 1$ states (this holds true in addition to the no-resetting argument).

\subsection{Phase diagram}
\label{sec:phase}

\begin{figure*}
    \centering
    \includegraphics[width=0.47\textwidth]{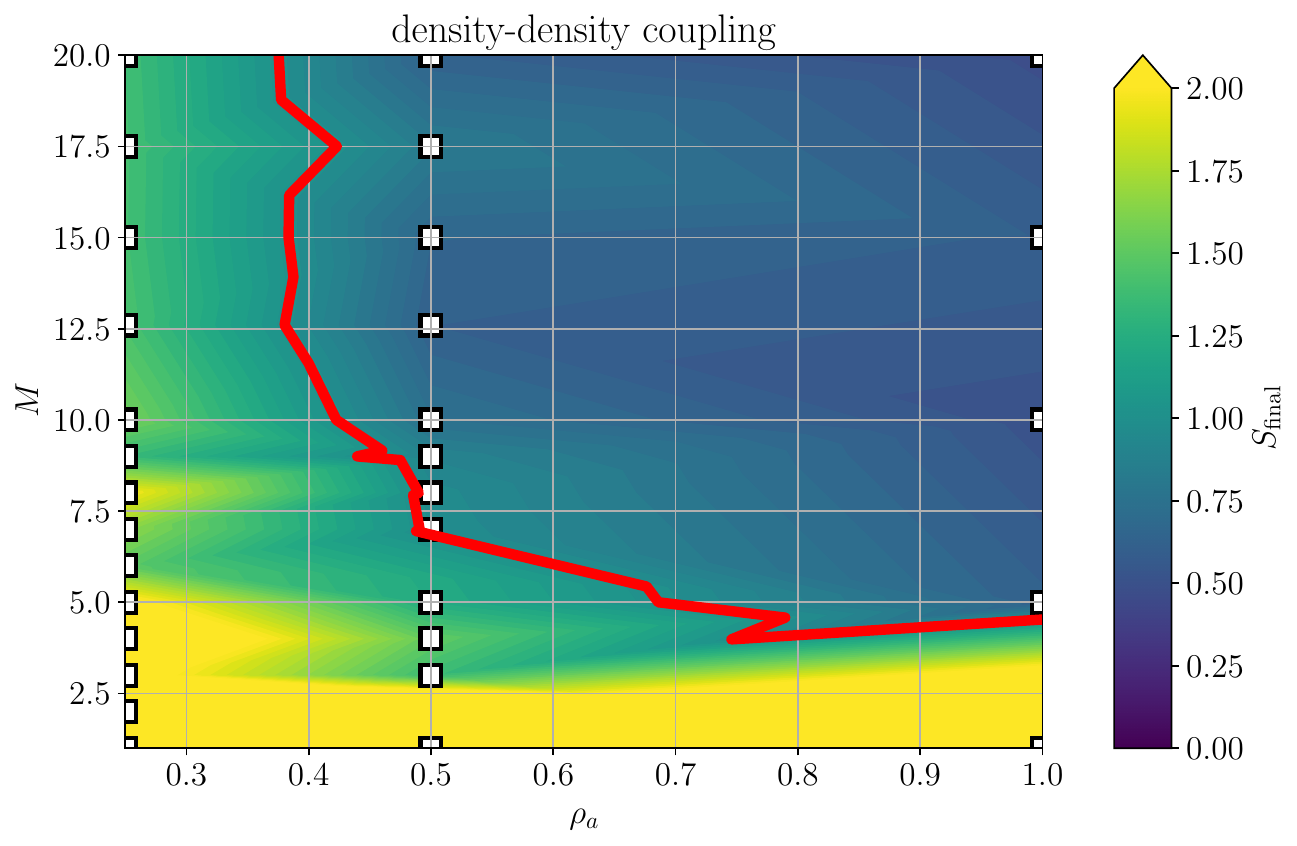}
    \hspace{0.02\textwidth}
    \includegraphics[width=0.46\textwidth]
    {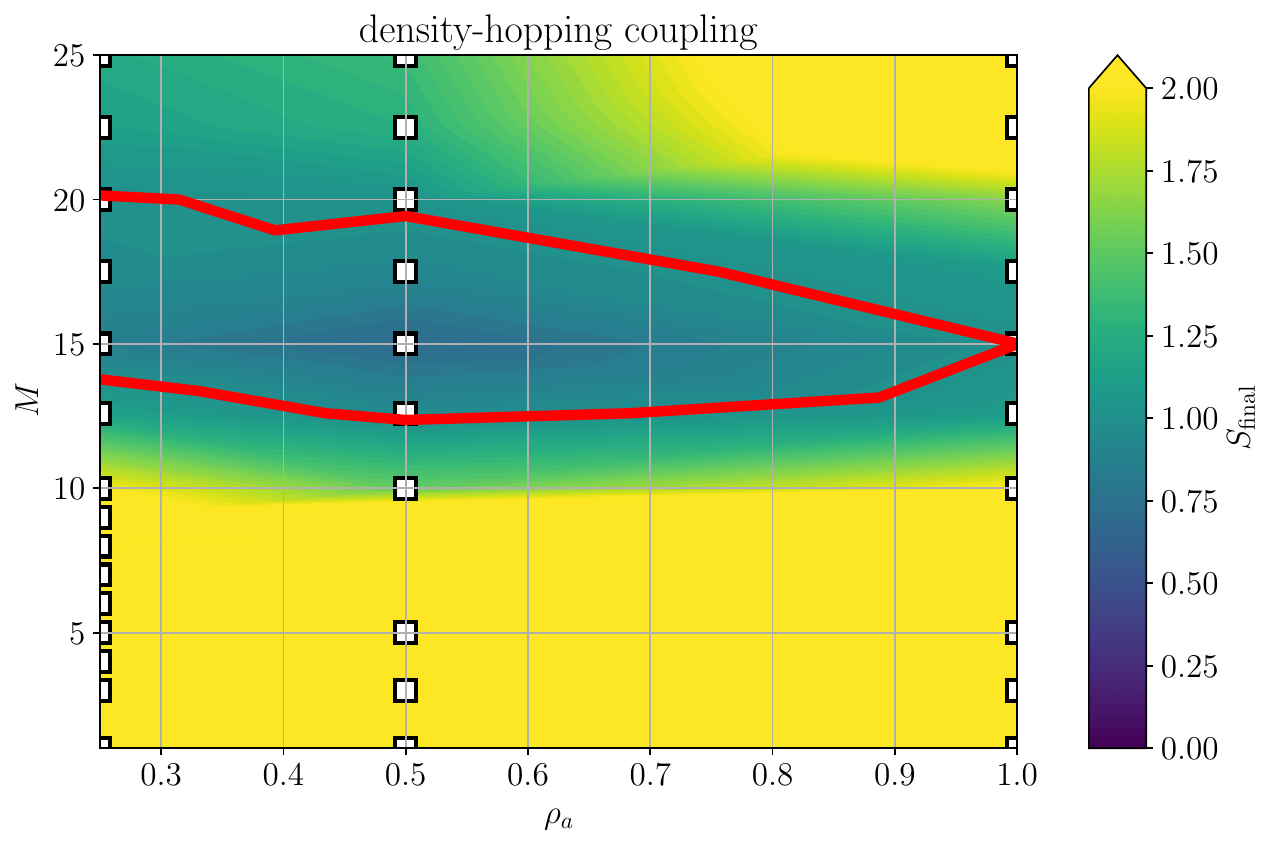}
    \caption{Qualitative phase diagrams for models (i) (left) and (ii) (right) in the parameter plane $(\rho_a, M)$ {\color{black} based on finite-size finite-time numerical data obtained in this work.} The color code shows the value $S(t=50)$ of the entanglement entropy at the end of time evolution  (dark blue: most disentangling behavior; yellow: most entangling). 
    The square symbols indicate the points for which MPS calculations were performed; the remainder is obtained by interpolation. The interpolation in between the data points is performed using the \texttt{tricontourf} routine \cite{matplotlib}. The red lines correspond to $S(t=50) = 1.0$ and serve as a
    rough estimate of the phase boundaries between the disentangling (area-law) and entangling (presumably, volume-law) phases. }
    \label{fig:phasediag_model1-2}
\end{figure*}

Based on the dynamics of the entropy, we are now in a position to investigate the phase diagrams for both models.
Figure~\ref{fig:phasediag_model1-2} shows the average entropy $S$ at the final time $t = 50$ of the simulation, for various choices of the ancilla concentration $\rho_a$ and measurement strength $M$ (the color-coding in the figure is obtained by interpolation between discrete points by using a routine specified in the figure caption). 
Of course, strictly speaking, this value does not automatically distinguish between the possible phases. However, inspecting the 
entropy curves in Figs.~\ref{fig:ent_model1} and \ref{fig:ent_model2}, we clearly see that large values of $S(t=50)$ correspond to unsaturated growth of $S$ with time, while low values are associated with decreasing entropy that tends to saturate at long times. With this indicator, in both panels of Fig.~\ref{fig:phasediag_model1-2}, we find a qualitative change between disentangling (blue regions) and entangling (yellow regions) types of behavior, with the border between them belonging to the green regions. 

This border is additionally visualized by a red line corresponding to $S(t=50)=1.0$ (this value is close to the value of the entropy in the initial state at $t=0$), 
which can be viewed as an estimate of the boundary between the disentangling and entangling types of behavior. In turn, based on the general hypothesis that the entanglement-entropy growth in {\color{black}generic} monitored interacting systems is a manifestation of the volume-law phase \footnote{{\color{black} 
The transition between the volume-law and area-law phases was established in the context of random quantum circuits, see Ref.~\cite{Fisher2022} for review. It is expected that it also holds in 
interacting Hamiltonian systems, see a discussion 
in Sec. VIII of Ref.~\cite{Poboiko2023a}. Analytical work in this direction is currently in progress.}},
this line also gives a rough estimate for the phase boundary between the area-law and volume-law phases.

For model (i), the increase of $M_c$ with increasing ancilla concentration $\rho_a$ is clearly visible. Moreover, the red line for this model goes almost vertically at $\rho_a\approx 0.4$ starting at $M=12.5$. This provides a hint at a possible divergence of $M_c$ at a finite ancilla concentration in this model.
The phase diagram for model (ii) has a more complex structure, reflecting the reentrance and weaker dependence on $\rho_a$ discussed above. This structure is somewhat reminiscent of the phase diagrams obtained for a single monitored qubit in Ref.~\cite{Poepperl2023}, where the same system-ancilla coupling was addressed (but with resetting of the ancilla pair after measurements).

We would like to reiterate that the phase diagrams obtained in this work and shown in
Fig.~\ref{fig:phasediag_model1-2}
are intended to demonstrate a clear qualitative difference in the entanglement dynamics for the two models, as well as the role of the detector concentration in each of them.
Obviously, they characterize the behavior of the systems at moderately large length scale and moderately large times, thus representing 
``finite-size finite-time phase diagrams''.
For a quantitative determination of the critical line separating distinct phases in the thermodynamic limit, much more involved computational efforts are required,  in combination with analytical input. 

At this point, it is worth recalling the situation 
for monitored 1D free fermions, where numerical analyses (naturally restricted to computationally accessible system sizes) suggested a phase transition between an area-law phase and a phase with an unsaturated entanglement.
It was, however, analytically demonstrated recently that, in the thermodynamic limit, a 1D {\color{black} free-fermion} system is in the area-law phase for an arbitrary strength (frequency) of measurements~\cite{Poboiko2023a}. For rare measurements, the system sizes required to detect the area law turn out to be exponentially large, which makes it extremely difficult to determine the thermodynamic-limit phase diagram by purely computational means.
 This emphasizes the necessity of an analytical background providing key input for solid quantitative computational studies of the transition. The present work provides important evidence regarding the role of the system-detector couplings and detector concentrations in the dynamics of ancilla-measured chains. We expect that this can be employed as a building ingredient for prospective theories of the transition in correlated systems.

\section{Summary and discussion}

\subsection{Summary}

In this paper, we have numerically modeled a quantum many-body lattice system (with a conserved particle number) coupled to a finite concentration of detectors (``ancillae'') that are subject to periodic measurements. The central questions that we addressed are how the dynamics of the entropy and of the density depend on the ancilla concentration $\rho_a$, the ancilla coupling $M$ to the main system, and on the type of this coupling. Specifically, we considered two types of coupling: model (i) with a density-density coupling, i.e., $S_z s_z$-type in the spin language, and model (ii) with a density-hopping coupling, i.e., $S_z s_x$-type, where {\boldmath$S$} and {\boldmath $s$} refer to the lattice site and to the detector coupled to it, respectively. In both models, the detectors are measured in the $s_z$ basis, so that the measured operator commutes with the coupling in the first model but does not commute in the second model. By using the MPS-based computational approach, we studied correlated chains of the length $L=40$ for $\rho_a=1$, $\rho_a = 1/2$, and $\rho_a =1/4$ (i.e., with $L$,  $L/2$, and $L/4$ detectors represented by ancilla pairs, respectively). 

We have found that, for model (i), the critical value $M_c$ of the measurement-induced entanglement transition is strongly dependent on the ancilla concentration $\rho_a$, see
Fig.~\ref{fig:phasediag_model1-2}.
Furthermore, our results indicate that $M_c$ diverges at some critical concentration $\rho_a^{(c)}$ (which is close to 1/4), so that 
for $\rho_a < \rho_a^{(c)}$ the system is in the entangling phase for any value of $M$.

For model (ii), the behavior is different in two key aspects. First, the dependence on $M$ is strongly non-monotonic, suggesting a re-entrance of the entangling phase at large $M$. Secondly, the system is much less sensitive to the concentration of detectors $\rho_a$ than in the case of model (i). 

We have complemented the analysis of the entanglement entropy with the particle density $n_j(t)$ in the chain. Importantly, we studied $n_j(t)$ for individual quantum trajectories, as the average of $n_j(t)$ over trajectories misses the physics related to the entanglement transition. 
For relatively weak couplings $M$, i.e., in the entangling phase, the density $n_i(t)$ for a given quantum trajectory fluctuates weakly around $n=1/2$. On the other hand, for large $M$, we observe clear long-living striped (``pajama'') patterns. This freezing of density suppresses the entanglement growth and is attributed to the QZVE.

\subsection{Outlook}

Let us conclude by briefly discussing prospects for future research. We expect that results of this work will be instrumental in boosting computational, analytical, and experimental studies in the directions outlined below.

An important open question in the physics of measurement-induced transitions is the effect of particle-number conservation. For monitored non-interacting fermions (with explicit particle-number conservation), the behavior of the entanglement entropy can be captured by the analysis of the particle-number  cumulants~\cite{Poboiko2023a}, with the entanglement transition in 2D systems coinciding with the transition for the density correlations~\cite{Poboiko2023b, Chahine2023}. At the same time, in a certain special class of random quantum circuits representing interacting systems (involving Haar-random gates and qudits with $d\to \infty$ states), a related ``charge sharpening'' transition was predicted to be distinct from the entanglement transition and to take place within the volume-law phase~\cite{Agrawal2022, Barratt2022}. It is thus a key open question whether the entanglement transition and the particle-number-fluctuation (or ``charge sharpening'') transitions coincide or are distinct for a realistic problem of interacting fermions (or hard-core bosons), like the one considered in the present work. In both cases, it is also important to understand how the entanglement and the density correlations influence each other (and, in particular, the corresponding scaling behavior), as well as whether the violation of the particle-number conservation could drastically affect the results obtained in the present work. 

Importantly, the above density correlations should be evaluated in a given quantum state (and only after this can be averaged over quantum trajectories), in similarity to density patterns for individual trajectories and to the distribution function $P(n;t)$ discussed in this paper (and also to the density clusterization that was observed for $\rho_a=1$ in Refs.~\cite{Doggen2022a} and \cite{Doggen2023a}). 
We expect that the MPS-based approach developed in this work may be extended to study quantitatively density fluctuations in models of monitored interacting 1D fermions or bosons and, in this way, to provide responses to the above questions from the computational perspective. It remains to be seen how universal the resolution of the above dichotomy (one vs. two transitions) is and, in particular, whether it may depend on the interaction strength, the type of the measurement protocol (like models (i) and (ii) in this paper) and on resetting (as in this paper) or non-resetting of detectors. Finally, it is interesting to study whether a commensurability of the ancilla periodicity with that of the main lattice is important in this context and whether random placement of ancillas would lead to any essential modifications.

\section{Acknowledgments}

We acknowledge collaboration with Y. Gefen and D.G. Polyakov on earlier related projects. We also thank D.G. Polyakov for useful discussions and comments on the manuscript. 
IVG acknowledges support from the Deutsche Forschungsgemeinschaft (DFG) via the grant GO 1405/6-1.

\textit{Note added:} After submission of the present work, a preprint appeared \cite{Cecile2024} that studies a problem of monitored interacting chains also utilizing an MPS-TDVP computational technique. This reference introduces an alternative approach (specific to MPS methods) to identifying measurement-induced transitions and distinguishing the entanglement transition from the charge-sharpening transition in interacting models. It will be interesting to apply this approach to ancilla-based measurements.

\begin{figure}[ht]
\includegraphics[width=0.97\columnwidth]{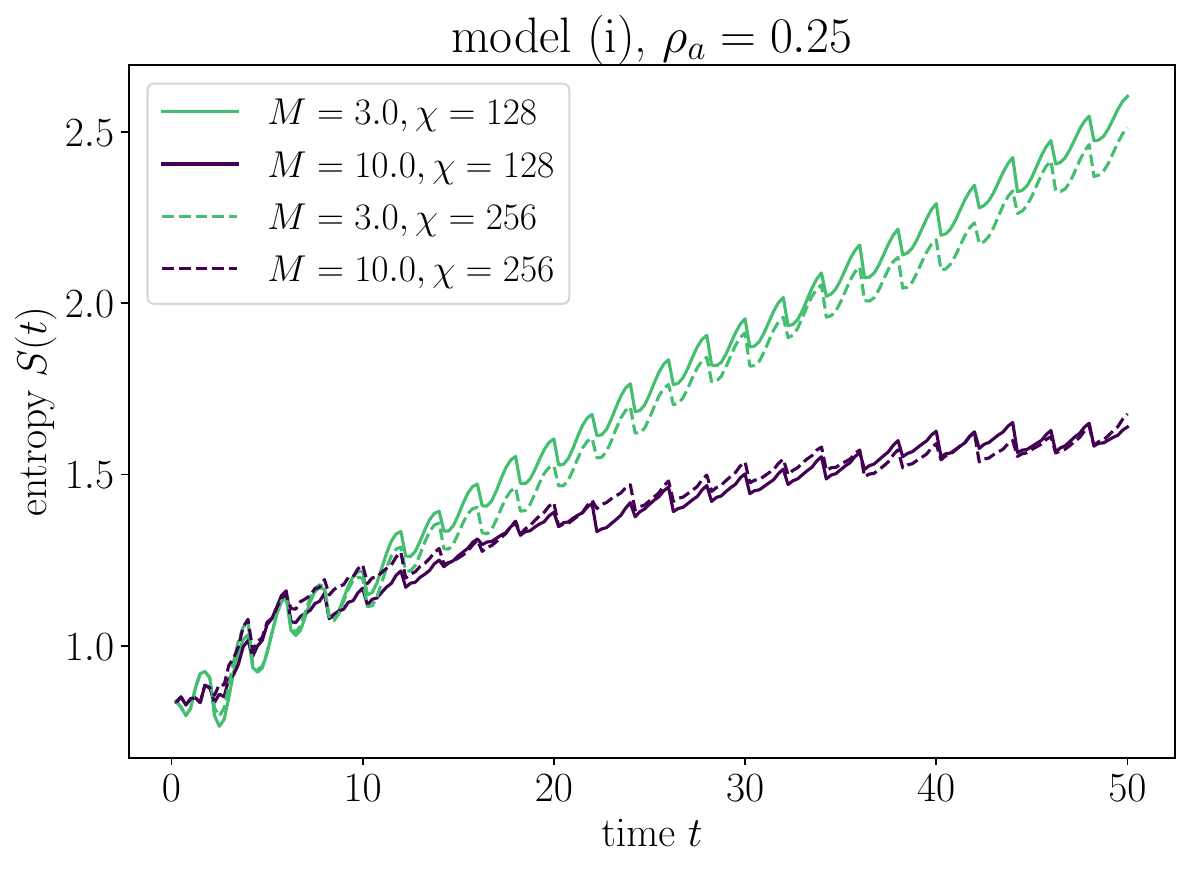} 
\vspace*{0.75cm}
\hspace*{0.1cm}
\includegraphics[width=0.95\columnwidth]{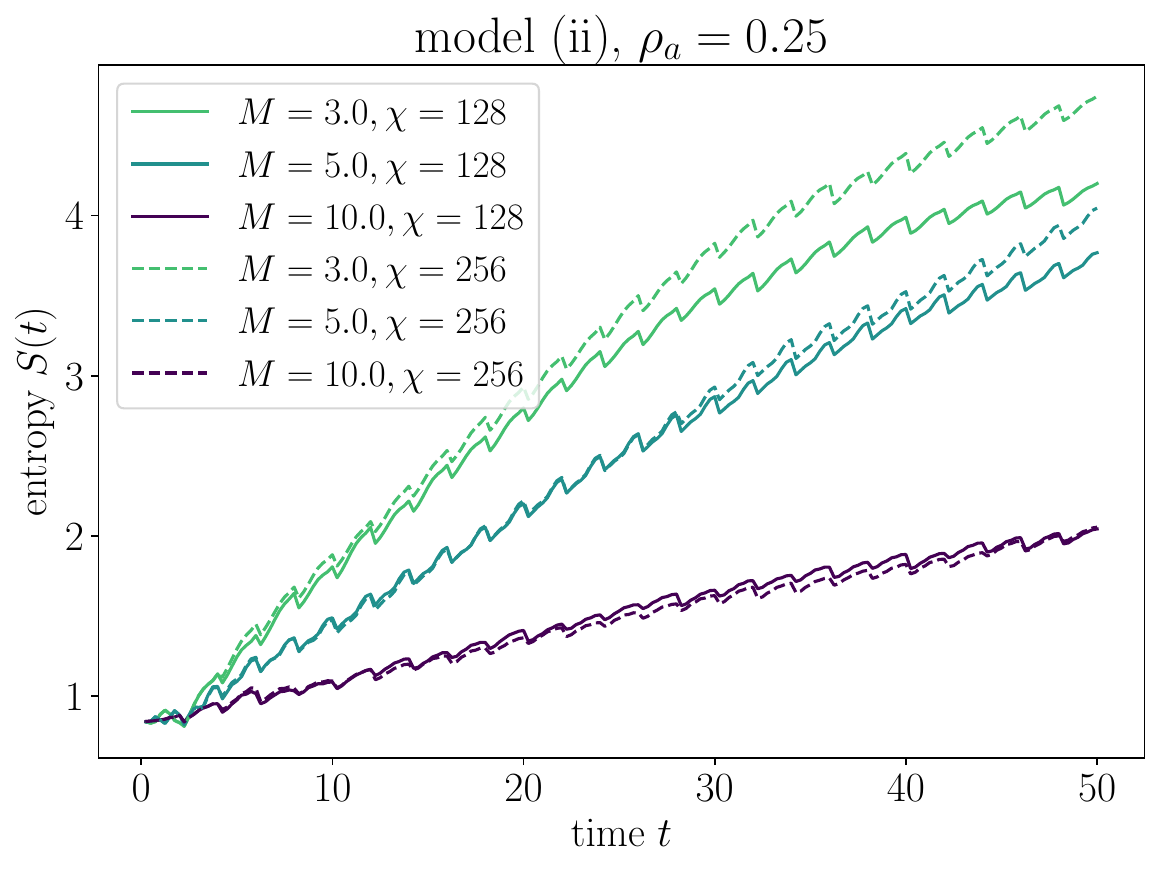}
    \caption{{\color{black} Time dependence of the entanglement entropy calculated with the two values of the MPS bond dimension: $\chi=128$ (as used in the main text; solid curves) and $\chi=256$ (dashed curves). Upper panel: model (i) at $\rho_a=1/4$. Lower panel: model (ii) at $\rho_a=1/4$. For $S(t)\lesssim 3$,
    the results for $\chi=128$ and $\chi=256$ are essentially identical (up to small statistical fluctuations related to a finite number of quantum trajectories), implying that the numerical cut-off set by $\chi=128$ does not affect the entropy curves.} }
    \label{fig:bench}
    \end{figure}

\color{black}

\appendix

\section{Benchmarking the role of the MPS bond dimension}
\label{appendix}

Here, we explicitly demonstrate that the bond dimension $\chi=128$ used in the main text is sufficient for the analysis of the entanglement behavior in both models addressed in this work. In Fig.~\ref{fig:bench}, we present the time-dependence of the entropy $S(t)$ for several parameter choices, including both models (i) and (ii), with the bond dimension $\chi=128$ (as in the main text) and with a twice larger bond dimension $\chi=256$.
The figure demonstrates the convergence of the entropy curves with bond dimension for $S(t)\lesssim 3$. (There are only small random deviations related to a finite size of the statistical ensembles. We do not enforce the measurement outcomes to be equal for different bond dimension, as the time evolution itself depends on the state of the system.)
Only for $S(t) \gtrsim 3$  systematic deviations of the $\chi=128$ curves down from the corresponding $\chi=256$ curves start to develop.

The range $S(t)\lesssim 3$ where the convergence up to $t=50$ is achieved covers all the values of parameters for model (i) used to construct the phase diagram in the left panel of Fig.~\ref{fig:phasediag_model1-2}, including the most entangling curve shown in Fig.~\ref{fig:ent_model1} ($\rho_a=1/4$ and $M=3$). For model (ii), the numerical cut-off imposed by the bond dimension affects the most entangling curves when $S(t)\gtrsim 3$, leading to a certain reduction of the calculated values of $S(t=50)$. 
These values, however, correspond anyway to the parameters located deeply inside the yellow regions [$S(t=50)>2$] in the right panel of Fig.~\ref{fig:phasediag_model1-2}, so that the numerical cut-off does not affect the obtained phase diagram.

\color{black}

\bibliography{ref}

\widetext

\renewcommand{\theequation}{S\arabic{equation}}
\renewcommand{\thefigure}{S\arabic{figure}}
\renewcommand{\thesection}{S\arabic{section}}
\setcounter{equation}{0}
\setcounter{figure}{0}

\vspace{1.5cm}

\begin{center}
{\Large\textbf{Supplemental Material to \\``Ancilla quantum measurements on interacting chains: Sensitivity of entanglement dynamics to the type and concentration of detectors''}}
\end{center}

\vspace{0.5cm}

In this Supplemental Material, we provide additional information on the results of numerical modeling for both models (i) and (ii), for the chain length $L=40$ and various values of the ancilla density $\rho_a$. Each figure corresponds to a point $(\rho_a, M)$ in the parameter space of the corresponding model, cf. {\color{black} Fig. 9} of the main text. In each figure, the top row shows the time evolution of average density in the chain (left), average entanglement entropy (middle), and average occupation of the ``red'' ancilla site (right), cf. Fig. 1 of the main text. The second (third) row shows analogous quantities for the quantum trajectory that is least (respectively, most) entangled at $t=50$.
The bottom panel displays the distribution function $P(n)$ of the local density in the main chain at times $t=0.25$, 25, and 50. For each model and $\rho_a$, the values of $M$ are chosen in such a way that the data illustrate the disentangling and entangling regions of the phase diagram,  {\color{black} Fig. 9} of the main text. 

\newpage

\begin{figure*}[!p]
    \centering
     \includegraphics[width=\textwidth]{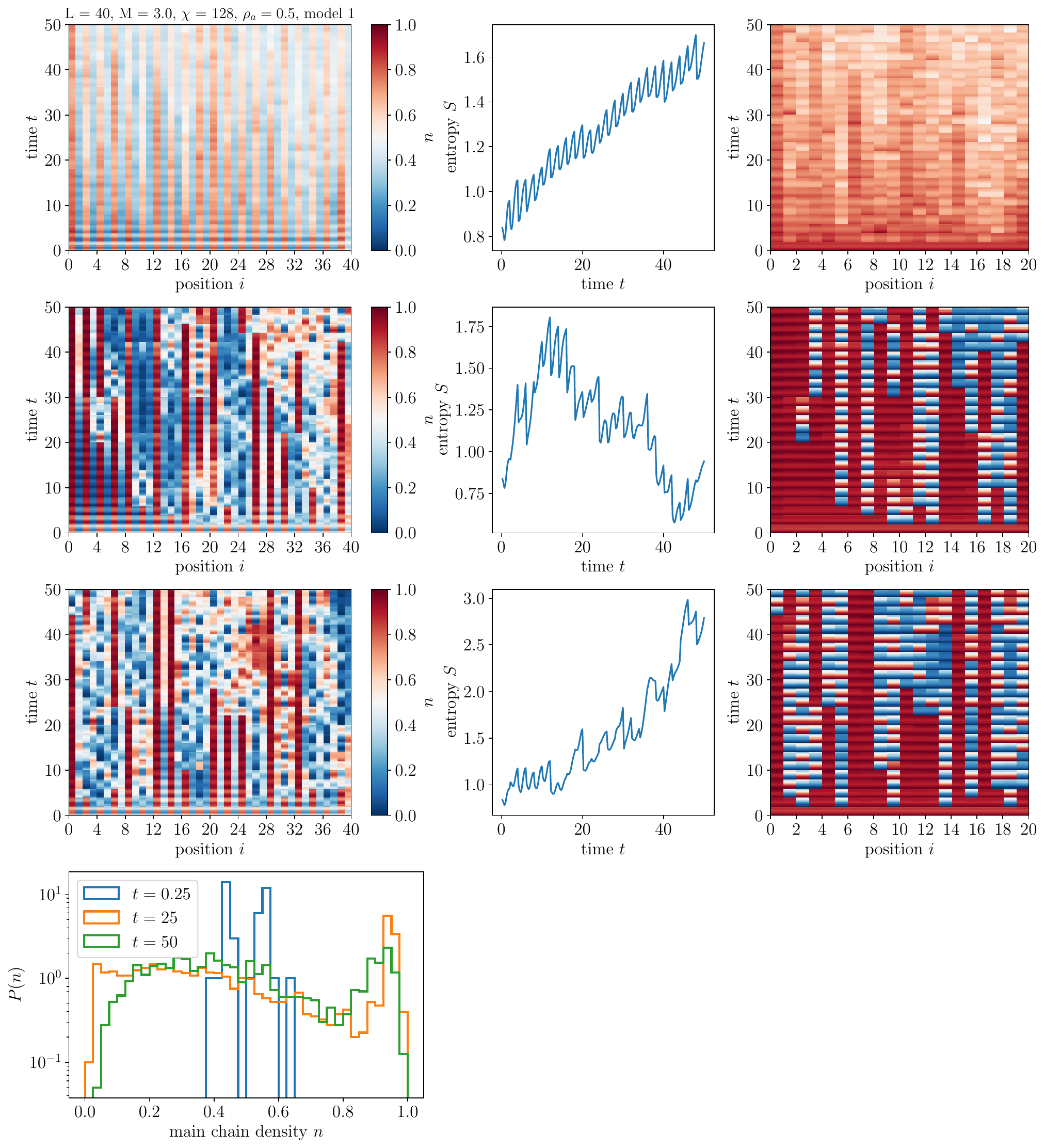}
 \caption{Model (i), $\rho_a = 1/2$, $M=3$.
 Top row:  time evolution of average density in the chain (left), average entanglement entropy (middle), and average occupation of the ``red'' ancilla site (right), cf. Fig. 1 of the main text. Second (third) row: analogous quantities for the quantum trajectory that is least (respectively, most) entangled at $t=50$.
Bottom panel: distribution function $P(n)$ of the local density in the main chain at times $t=0.25$, 25, and 50.
 }
 \label{figS1}
\end{figure*}

\begin{figure*}[!p]
    \centering
    \includegraphics[width=\textwidth]{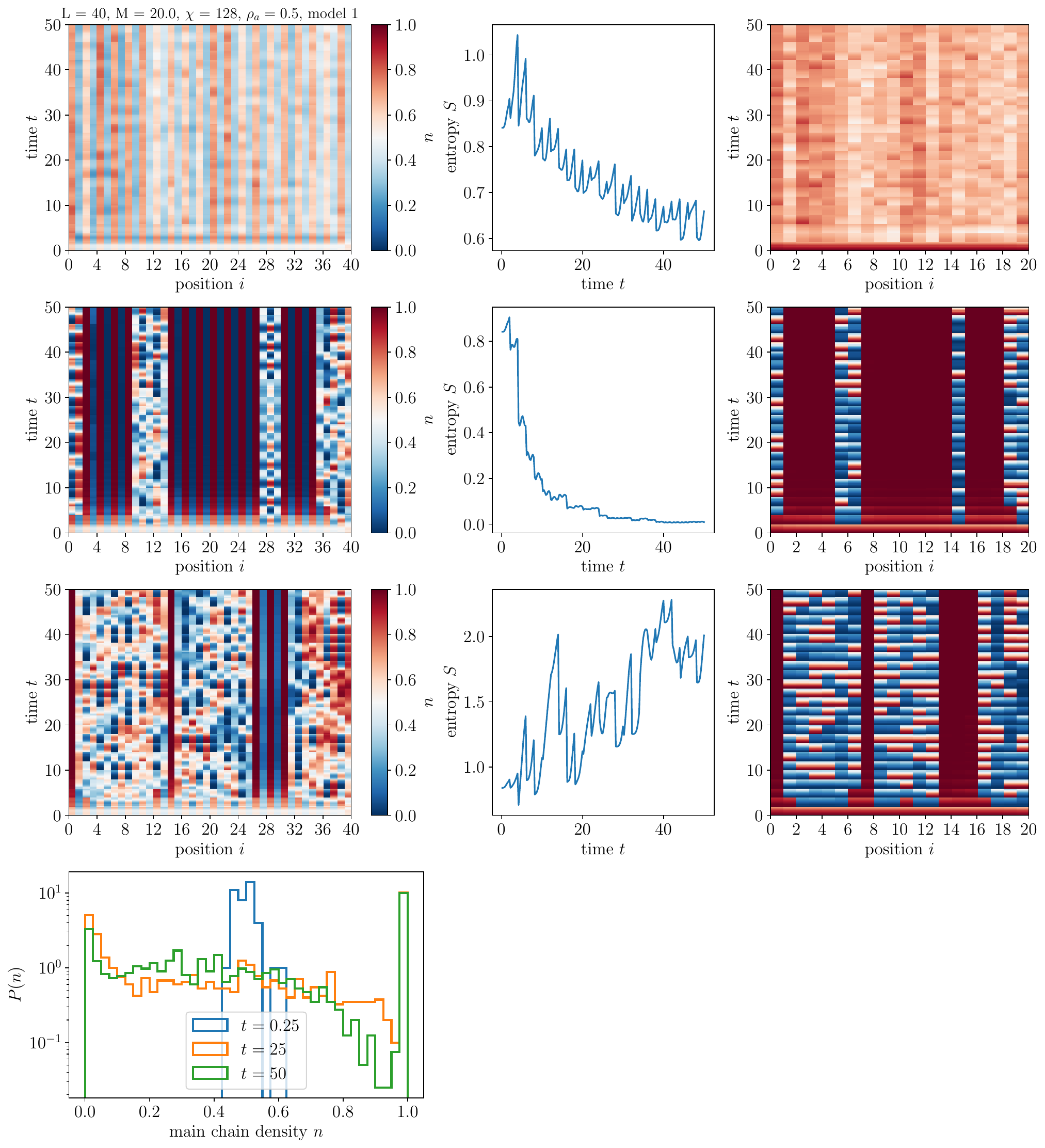}
    \caption{Same as Fig.~\ref{figS1} for model (i), $\rho_a = 1/2$, $M=20$.}
   \end{figure*}

\begin{figure*}[!p]
    \centering
       \includegraphics[width=\textwidth]{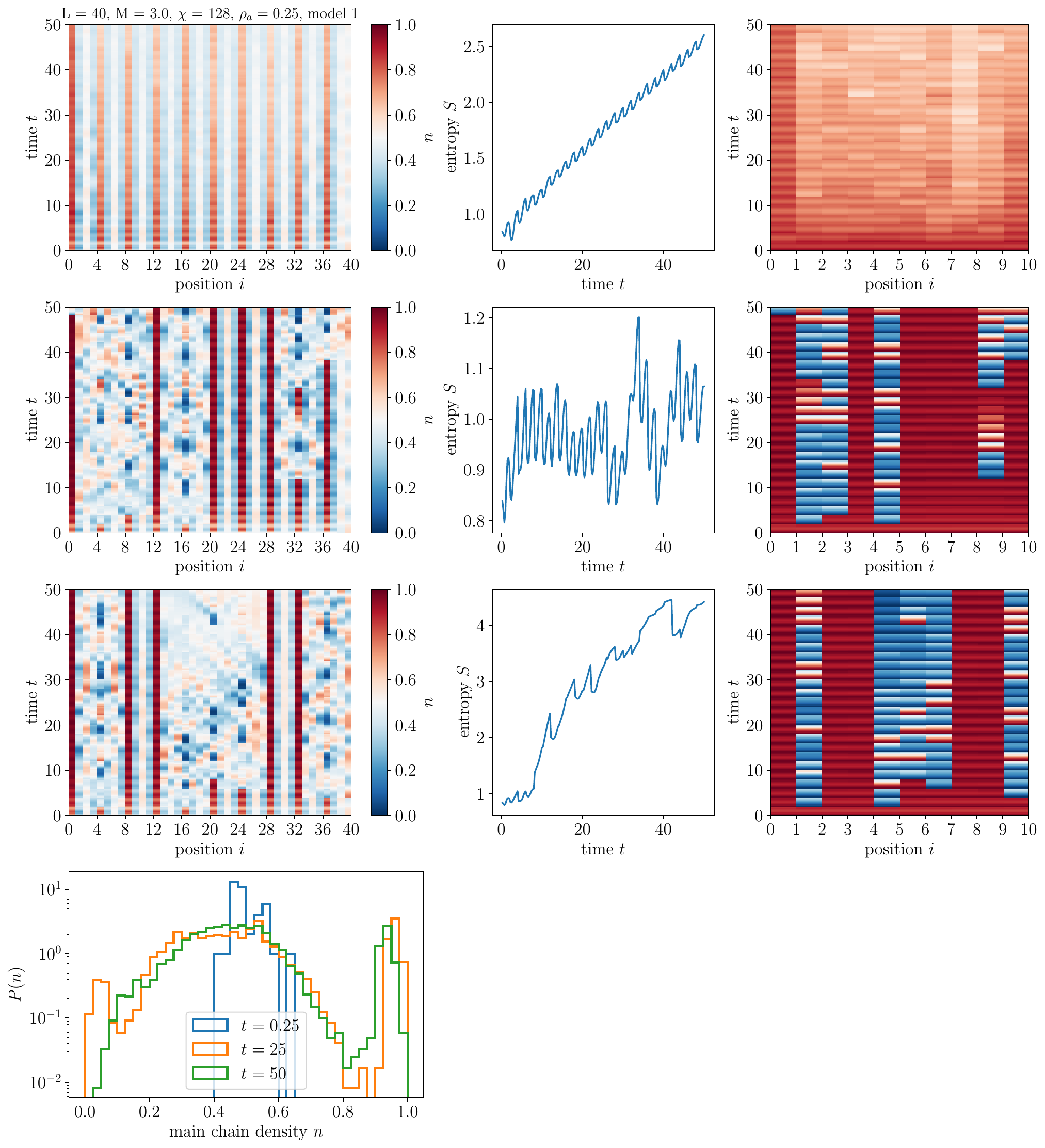}
       \caption{Same as Fig.~\ref{figS1} for model (i), $\rho_a = 1/4$, $M=3$.}
 \end{figure*}

\begin{figure*}[!p]
    \centering
    \includegraphics[width=\textwidth]{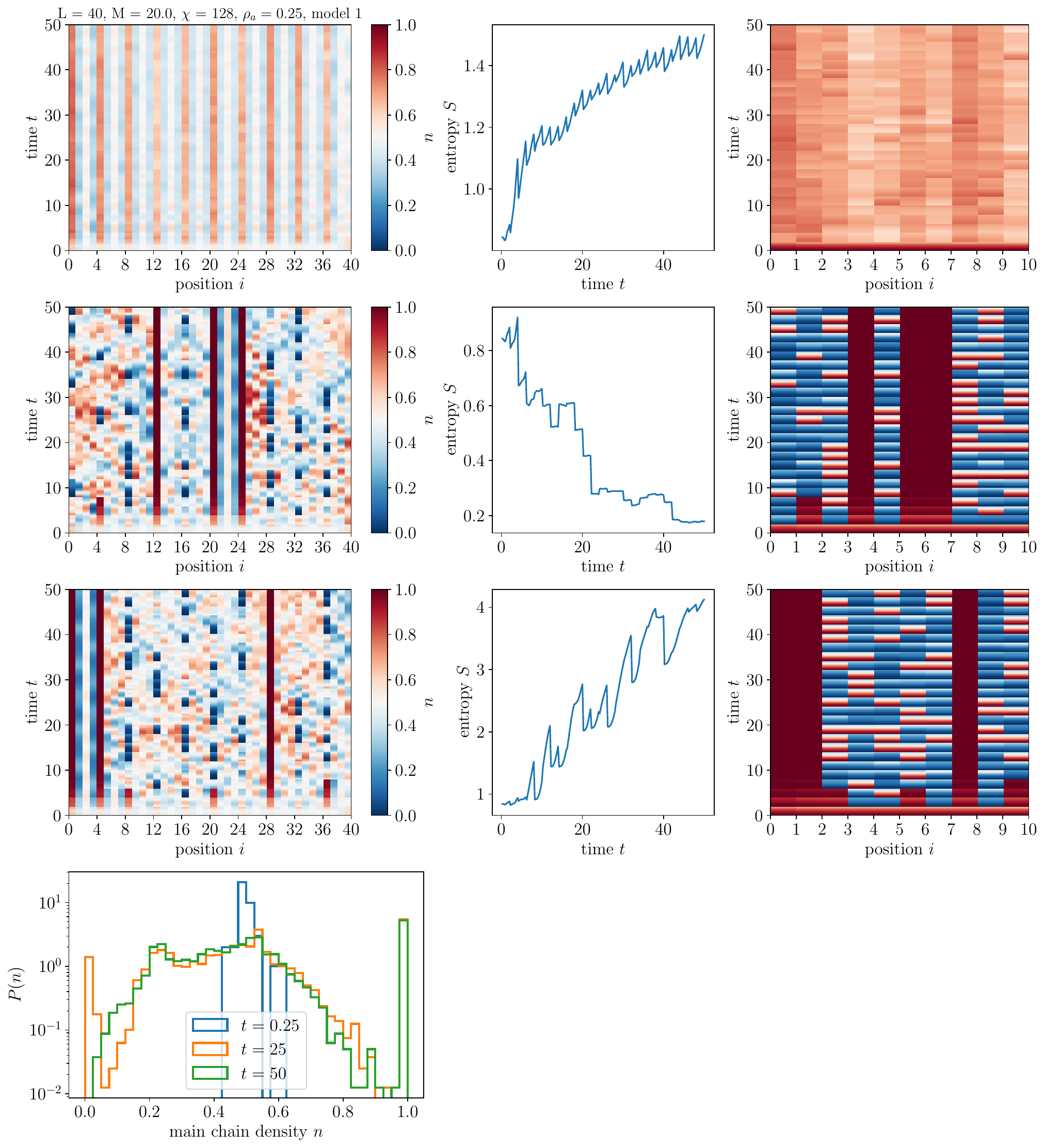}
      \caption{Same as Fig.~\ref{figS1} for model (i), $\rho_a = 1/4$, $M=20$.}
\end{figure*}


\begin{figure*}[!p]
    \centering
    \includegraphics[width=\textwidth]{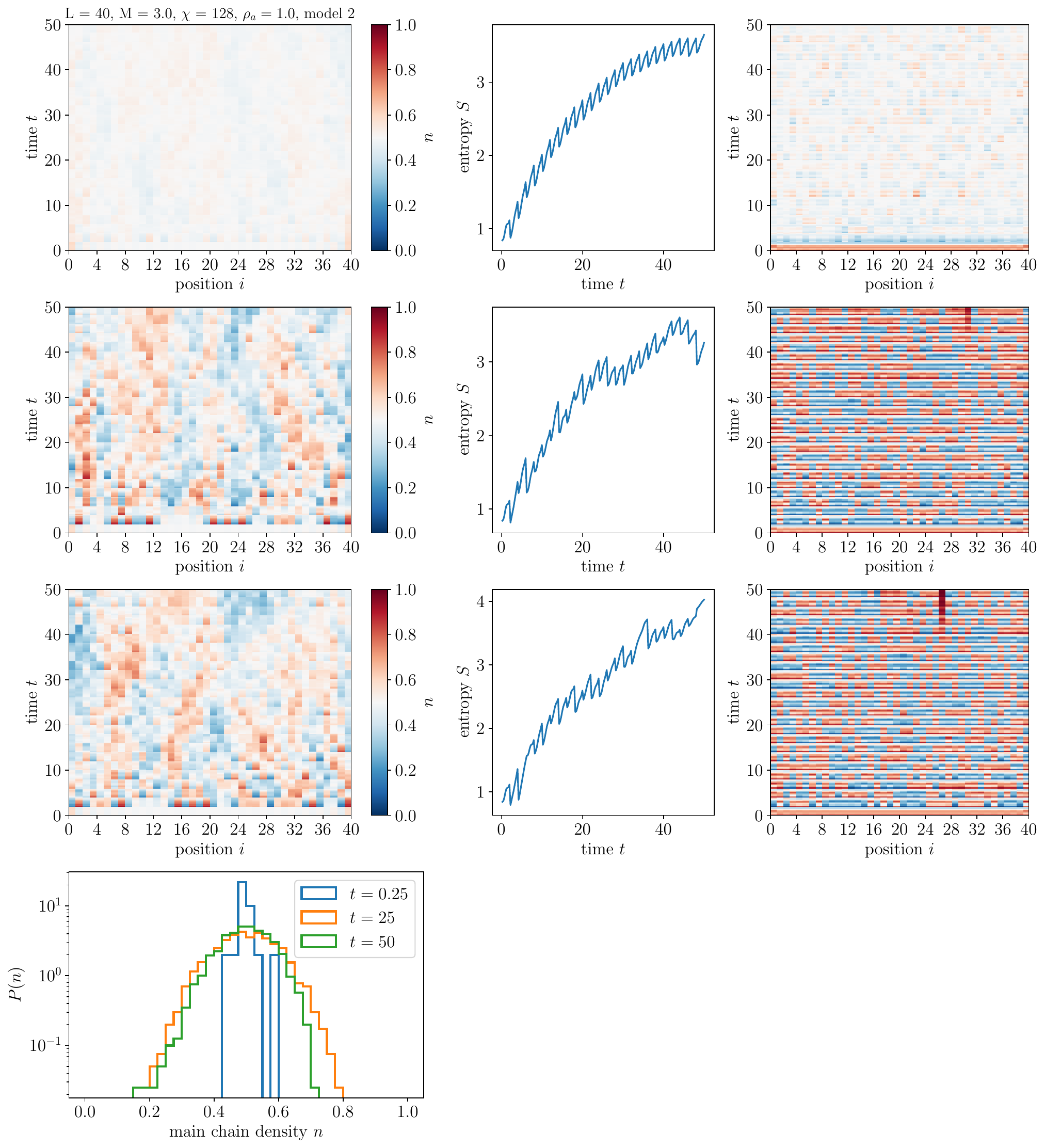}
\caption{Same as Fig.~\ref{figS1} for model (ii), $\rho_a = 1$, $M=3$.}
 \end{figure*}

\begin{figure*}[!p]
    \centering
    \includegraphics[width=\textwidth]{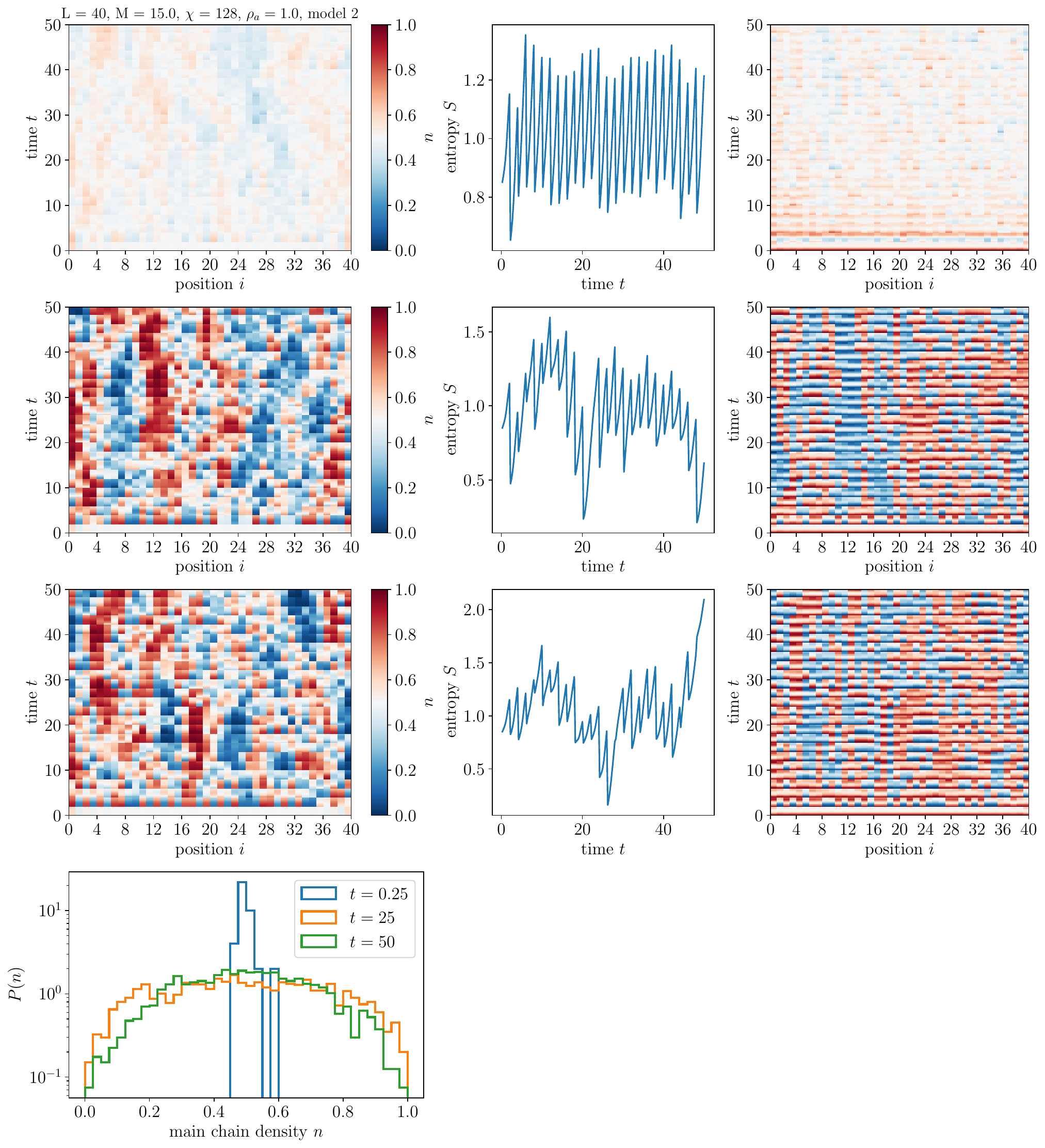}
\caption{Same as Fig.~\ref{figS1} for model (ii), $\rho_a = 1$, $M=15$.}
\end{figure*}

\begin{figure*}[!p]
    \centering
    \includegraphics[width=\textwidth]{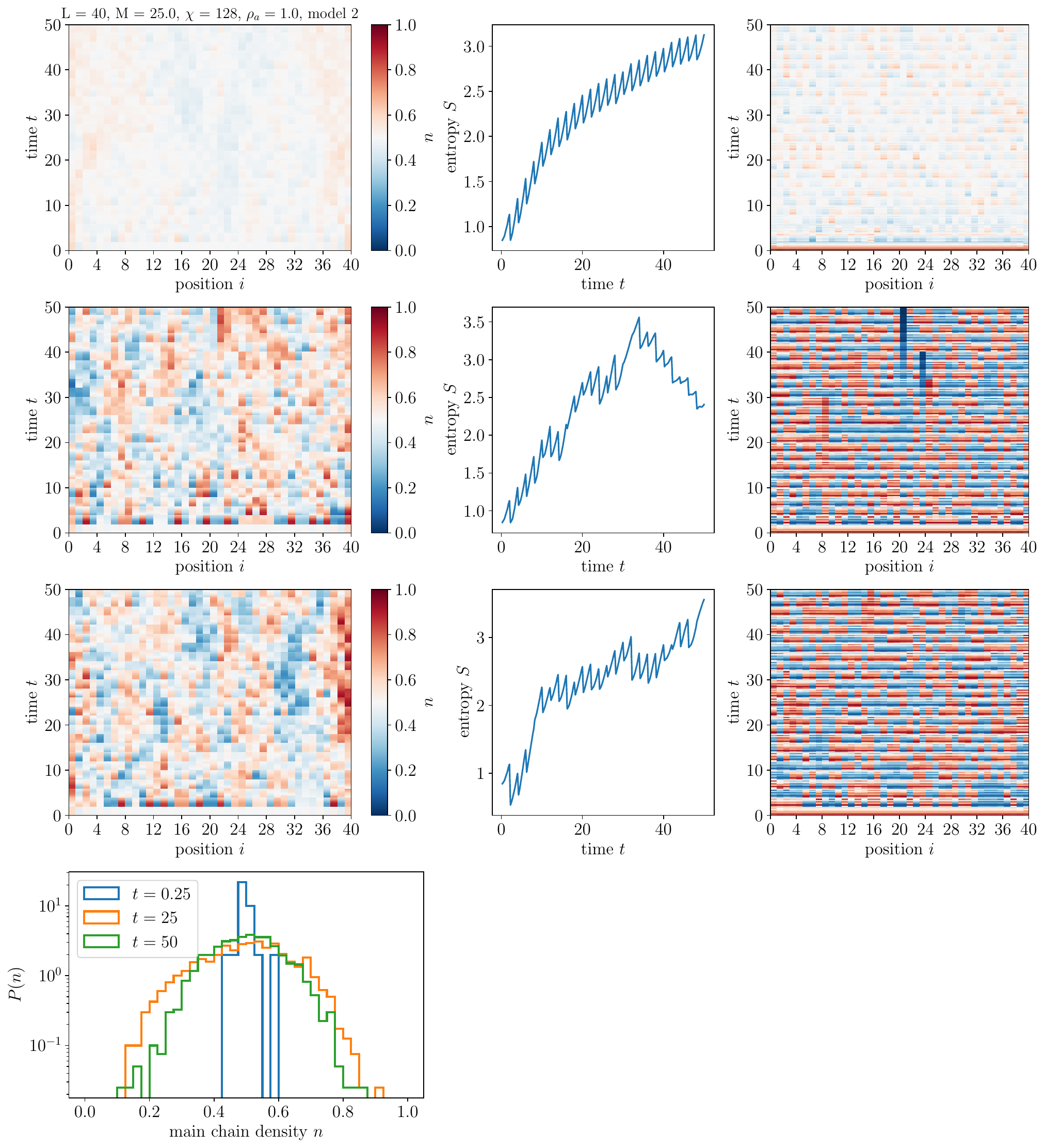}
    \caption{Same as Fig.~\ref{figS1} for model (ii), $\rho_a = 1$, $M=25$.}
\end{figure*}

\begin{figure*}[!p]
    \centering
    \includegraphics[width=\textwidth]{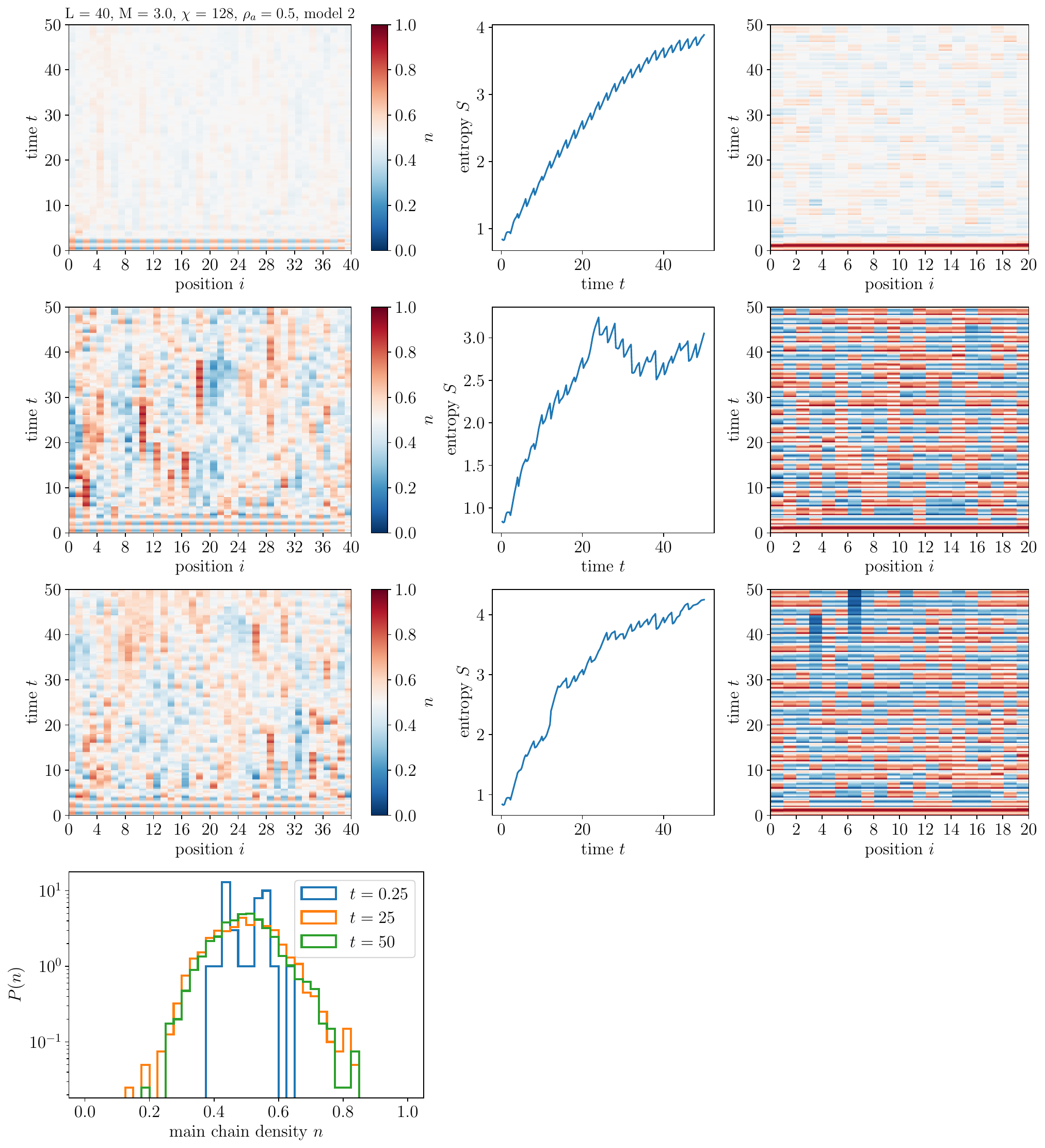}
    \caption{Same as Fig.~\ref{figS1} for model (ii), $\rho_a = 1/2$, $M=3$.}
\end{figure*}

\begin{figure*}[!p]
    \centering
        \includegraphics[width=\textwidth]{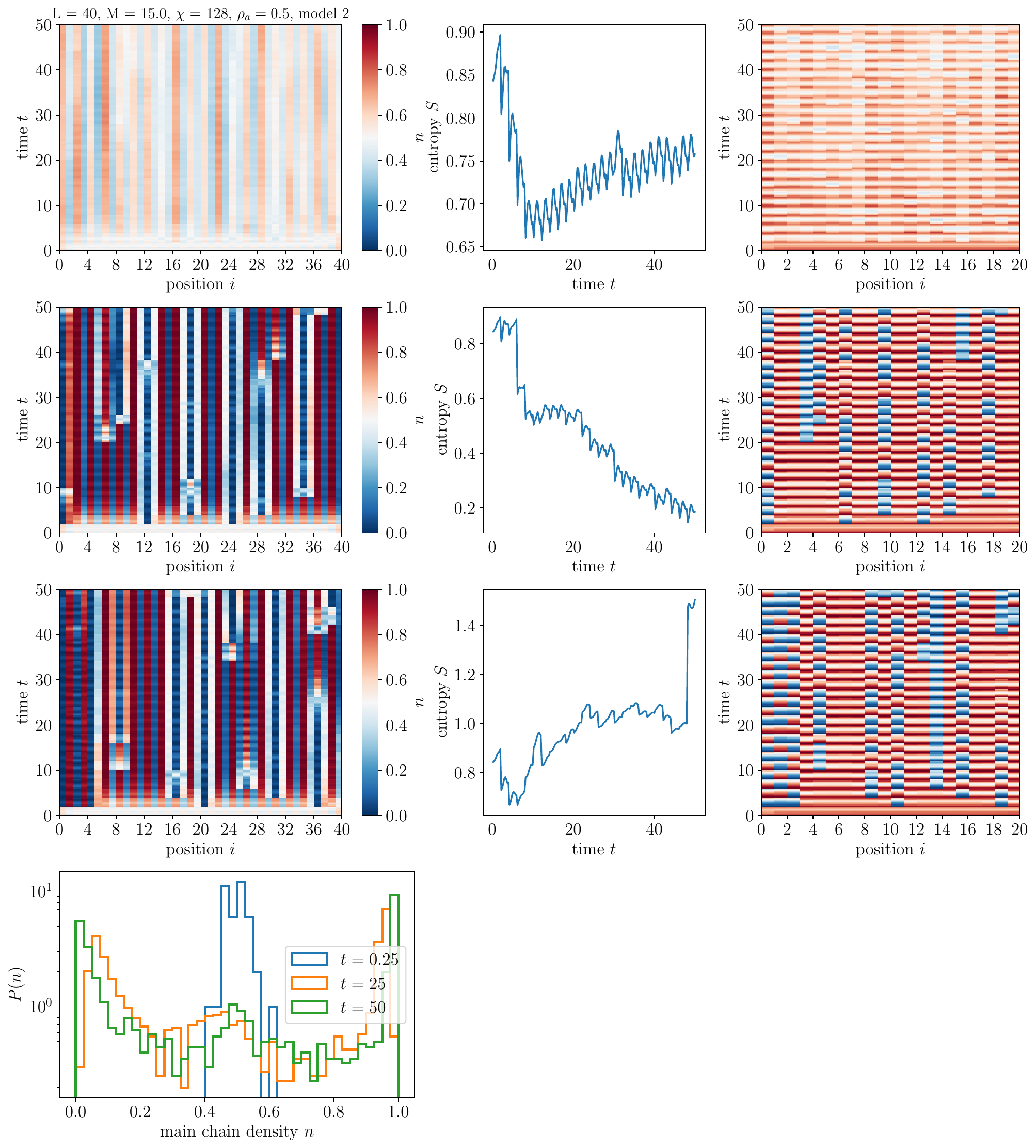}
     \caption{Same as Fig.~\ref{figS1} for model (ii), $\rho_a = 1/2$, $M=15$.}
 \end{figure*}

\begin{figure*}[!p]
    \centering
    \includegraphics[width=\textwidth]{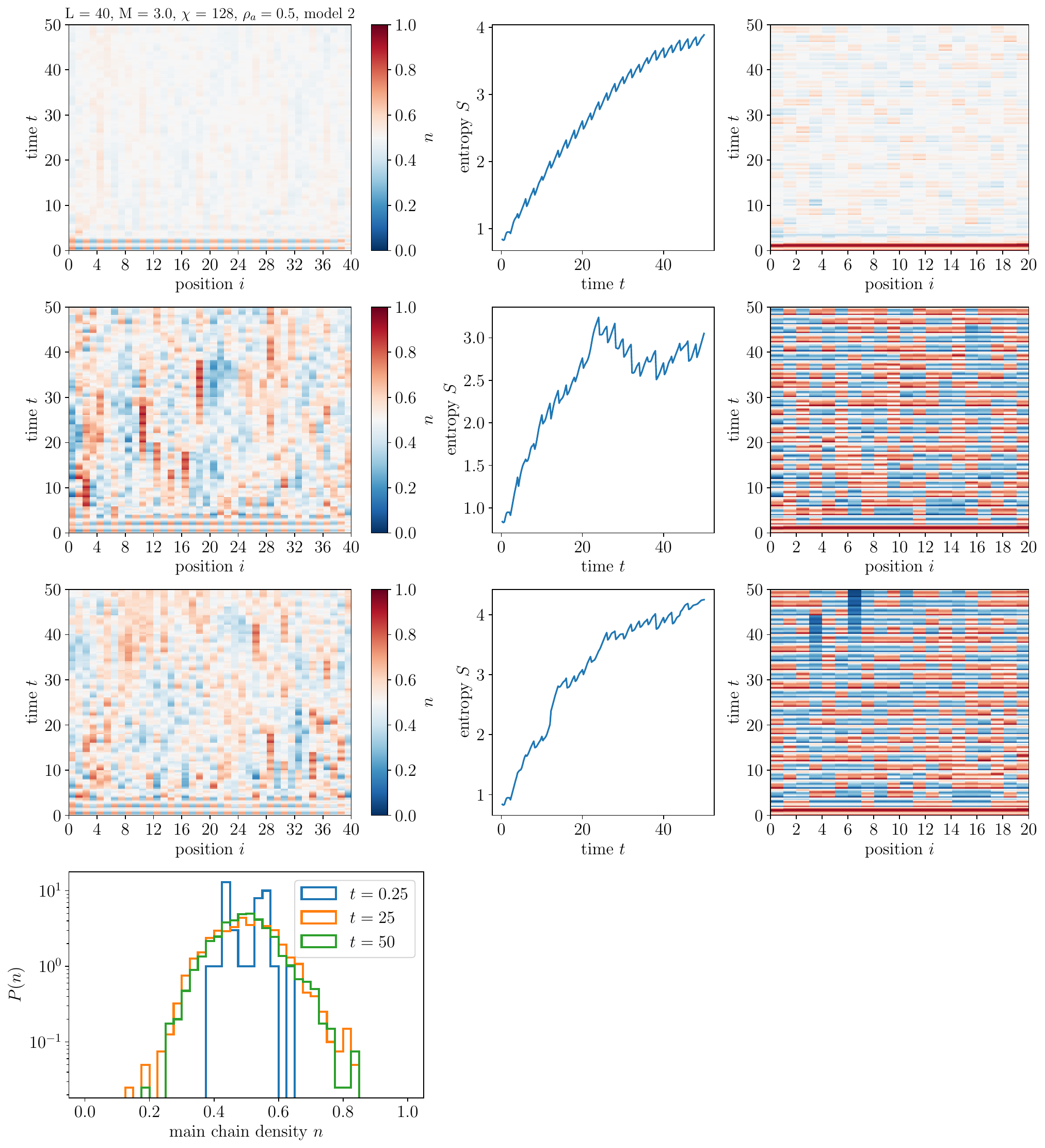}
 \caption{Same as Fig.~\ref{figS1} for model (ii), $\rho_a = 1/4$, $M=3$.}
 \end{figure*}

\begin{figure*}[!p]
    \centering
    \includegraphics[width=\textwidth]{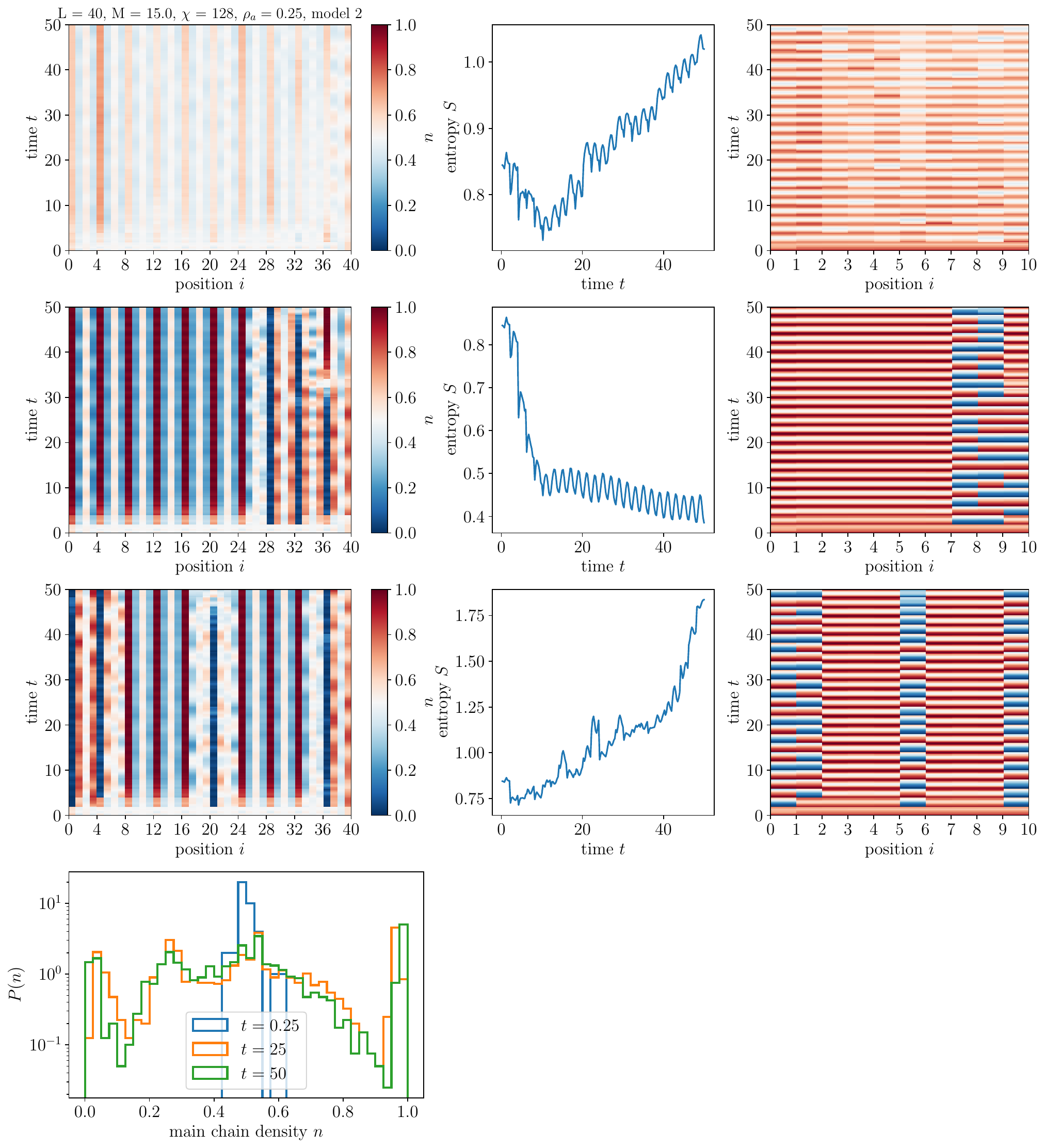}
     \caption{Same as Fig.~\ref{figS1} for model (ii), $\rho_a = 1/4$, $M=15$.}
\end{figure*}

\end{document}